\documentstyle[12pt]{article}
\parskip=\medskipamount                 % space between paragraphs
\thicklines                     % thick straight lines for pictures
\def\ordless{{\lower2mm\hbox{$\,\stackrel{\textstyle <}{\sim}\, $}}}
\def\ordgt{{\lower2mm\hbox{$\,\stackrel{\textstyle >}{\sim}\, $}}}

\newcommand{\bim}[6]{\bibitem{#1}#2, {\em #3\/}$\;${\bf
#4}$\;$(#5)$\;${#6}.}
\catcode`@=11
\@addtoreset{equation}{section}
\catcode`@=12

\begin{document}
\begin{titlepage}
\centerline{\hfill                 UTTG-06-93}
\centerline{\hfill                 hep-th/9303099}
\vfill
\begin{center}
{\large \bf A Large $k$ Asymptotics of Witten's Invariant of Seifert
Manifolds}
\\
\bigskip \centerline{L. Rozansky\footnote{Work supported by NSF Grant
9009850 and R. A. Welch Foundation.}}

\centerline{\em Theory Group, Department of Physics, University of
Texas at Austin}
\centerline{\em Austin, TX 78712-1081, U.S.A.}

\vfill
{\bf Abstract}

\end{center}
\begin{quotation}
We calculate a large $k$ asymptotic expansion of the exact surgery
formula for Witten's $SU(2)$ invariant of Seifert manifolds. The
contributions of all flat connections are identified. An agreement
with the 1-loop formula is checked. A contribution of the irreducible
connections appears to contain only a finite number of terms in the
asymptotic series. A 2-loop correction to the contribution of the
trivial connection is found to be proportional to Casson's invariant.

\end{quotation} \vfill
\end{titlepage}
\pagebreak

\tableofcontents
\pagebreak

%+++++++++++++++++++++++++++++++++++++++++++++++++++
\section{Introduction}
%+++++++++++++++++++++++++++++++++++++++++++++++++++

A Chern-Simons action is an ``almost'' gauge invariant function of a
gauge connection on a 3-dimensional manifold ${\cal M}$:
\begin{equation}
S_{CS}=\frac{1}{2\pi}\epsilon^{\mu\nu\rho}{\rm Tr}
\int_{\cal M}(A_{\mu}\partial_{\nu}A_{\rho}+\frac{2}{3}A_{\mu}A_{\nu}
A_{\rho})d^{3}x
\label{1}
\end{equation}
(a trace is taken in the fundamental representation of the gauge group
$G$). A quantum field theory built upon this action is topological.
This means that a partition function presented by a path integral over
the gauge equivalence classes of connections
\begin{equation}
Z({\cal M},k)=\int{\cal D}A_{\mu}e^{ikS_{CS}[A_{\mu}]}
\label{2}
\end{equation}
does not depend on the metric of the manifold ${\cal M}$ and is
therefore its topological invariant. An exact calculation of this
invariant was carried out by Witten in his paper \cite{W} on Jones
polynomial. The calculation requires a construction of ${\cal M}$ by a
surgery on a link in $S^{3}$ (or in some other simple manifold, say
$S^{1}\times S^{2}$). Reshetikhin and Turaev proved in \cite{RT} that
Witten's procedure really leads to a topological invariant. Their
proof does not use path integral (\ref{2}) which is not a rigorously
defined object for mathematicians yet.

A Chern-Simons action enters the exponential of the path integral
(\ref{2}) with an arbitrary integer factor $k$. Its inverse $k^{-1}$
plays a role of the Planck constant $\hbar$, which appears in quantum
theories and sets a scale of quantum effects. A stationary phase
approximation for the integral (\ref{2}) in the limit of $k\rightarrow
\infty$ expresses a partition function $Z({\cal M},k)$ as an
asymptotic series in $k^{-1}$. Physicists call this series a ``loop
expansion'', because the terms of order $k^{1-n}$ come from the
$n$-loop Feynman diagrams.

The loop expansion of $Z({\cal M},k)$ has been studied in \cite{AS},
\cite{ALR} and \cite{GMM}, as well as in papers \cite{DBN} and
\cite{K} which were aimed at producing Vassiliev's knot invariants.
Feynman rules were formulated, however the actual calculation of loop
corrections for particular manifolds went only up to the 1-loop order.
The 1-loop correction was found in \cite{W},\cite{FG},\cite{J} to
contain such invariants of the manifold ${\cal M}$ as the
Reidemeister-Ray-Singer torsion, spectral flow and dimensions of
cohomologies.

Thus there are two different methods of calculating $Z({\cal M},k)$: a
``surgery calculus'' of Witten-Reshetikhin-Turaev and a loop expansion
which is a standard method of quantum field theory. Both methods
should give the same value of $Z({\cal M},k)$ if the path integral
(\ref{2}) has the properties that physicists expect it to have.
D. Freed and R. Gompf suggested to check this by computing an exact
value of $Z({\cal M},k)$ for large values of $k$ through the surgery
calculus and then comparing it to the quantum field theory 1-loop
approximation. They carried out their program in \cite{FG} for some
lens spaces and Brieskorn spheres. A computer calculation showed a
close correspondence between the values of exact and 1-loop partition
functions. In a subsequent paper \cite{J}, L. Jeffrey used a Poisson
resummation trick to derive analyticly a large $k$ expansion of an
exact surgery formula for lens spaces and mapping tori. She also
observed a correspondence between the surgical and 1-loop expressions
(at least up to some minor factors, which we will discuss in the next
section). Similar results were obtained in \cite{G}.

In this paper we carry out a large $k$ expansion of an exact
surgery formula for Seifert manifold $SU(2)$ invariants. We identify
the contributions of all flat connections and show that they
correspond to the slightly modified 1-loop approximation formula of
\cite{FG} and \cite{J}. In contrast to the lens spaces, Seifert
manifolds have irreducible flat connections. We find rather
surprisingly that although the reducible connections contribute to
all orders in loop expansion, the contribution of irreducible
connections appers to be finite loop exact. We also find that a 2-loop
correction to the contribution of the trivial connection is
proportional to Casson's invariant.

In section 2 we review the basic features of loop expansion and
surgery calculus. Section 3 describes an application of the Poisson
resummation to the surgery formula for Seifert manifolds with 3
fibers. In section 4 the asymptotic expansion of the surgery formula
for those manifolds is compared to the 1-loop formula. In section 5
a Poisson resummation is applied to a general $n$-fibered Seifert
manifold and the contributions of both irreducible and reducible flat
connections are calculated.

\noindent
\underline{Summary of Results}
\nopagebreak

For reader's convenience we summarize briefly the main results of
the calculations in Section~\ref{*3}. We present the $SU(2)$ Witten's
invariant of a 3-fibered Seifert manifold
$X(\frac{p_{1}}{q_{1}},\frac{p_{2}}{q_{2}},\frac{p_{3}}{q_{3}})$ as a
sum over flat connections in the spirit of eqs.~(\ref{2.3}) and
(\ref{2.4}).

The contribution of an irreducible connection is 2-loop exact:
\begin{eqnarray}
&\frac{1}{2}\exp\left[2\pi iK\sum_{i=1}^{3}
\left(\frac{r_{i}}{p_{i}}\tilde{n}_{i}^{2}-q_{i}s_{i}\lambda^{2}
\right)\right]e^{2\pi i\lambda+i\frac{3\pi}{4}{\rm sign}
\left(\frac{H}{P}\right)}{\rm sign}(P)&
\nonumber\\
&\times
\left[\prod_{i=1}^{3}\frac{1}{\sqrt{|p_{i}|}}
2i\sin\;2\pi\left(\frac{r_{i}}{p_{i}}\tilde{n}_{i}+s_{i}\lambda
\right)\right]e^{-\frac{i\pi}{2K}\phi}&.
\label{S.1}
\end{eqnarray}
Here
\begin{eqnarray}
&
P=p_{1}p_{2}p_{3},\;H=p_{1}p_{2}q_{3}+p_{1}q_{2}p_{3}+
q_{1}p_{2}p_{3},\;\lambda=0,\frac{1}{2},&
\nonumber\\
&\tilde{n}_{i}=n_{i}+q_{i}\lambda,\;n_{i}\in{\bf Z},\;\;
ps-qr=1,\;s,r\in{\bf Z},&
\label{S.01}
\end{eqnarray}
$s(q,p)$ is a Dedekind sum. For more details see
subsection~(\ref{*4.1}).
The phase
\begin{equation}
\phi=3{\rm sign}\left(\frac{H}{P}\right)+
\sum_{i=1}^{3}\left(12s(q_{i},p_{i})-\frac{q_{i}}{p_{i}}\right)
\label{S.2}
\end{equation}
is the 2-loop correction. As we will see, this phase appears in the
contributions of all flat connections.

The contribution of a reducible connection contains an asymptotic
series of loop corrections:
\begin{eqnarray}
&-\exp\;2\pi
iK\left[\sum_{i=1}^{3}\frac{r_{i}}{p_{i}}n_{i}^{2}+
\frac{H}{P}c_{0}^{2}\right]\frac{e^{i\frac{\pi}{2}{\rm sign}
\left(\frac{H}{P}\right)}}{\sqrt{2K|H|}}{\rm
sign}(P)e^{-\frac{i\pi}{2K}\phi}&
\nonumber\\
&\times \sum_{j=0}^{\infty}\frac{1}{j!}(8\pi iK)^{-j}
\left(\frac{P}{H}\right)^{j}
\left.\left[\partial_{c}^{(2j)}\frac{\prod_{i=1}^{3}
2i\sin\left(2\pi\frac{r_{i}n_{i}+c}{p_{i}}\right)}{2i\sin 2\pi c}
\right]\right|_{c=c_{0}}&
\label{S.3}
\end{eqnarray}
here $c_{0}=\frac{P}{H}\sum_{i=1}^{3}\frac{n_{i}}{p_{i}}$,
for notations see section~\ref{*4.2}. A logarithm of this series has
to be calculated in order to put (\ref{S.3}) into the
form~(\ref{2.4}).

A reducible flat connection for which $c_{0}=0,\;1/2$
(see ``point on a face'' in subsection~\ref{*4.3}), contributes
\begin{eqnarray}
&-\exp\;2\pi iK\left[\sum_{i=1}^{3}\frac{r_{i}}{p_{i}}n_{i}^{2}+
\frac{H}{P}c_{0}^{2}\right]\frac{e^{i\frac{\pi}{2}{\rm sign}
\left(\frac{H}{P}\right)}}{\sqrt{2K|H|}}e^{2\pi ic_{0}}
{\rm sign}(P)e^{-\frac{i\pi}{2K}\phi}&
\nonumber\\
&\times
\left\{e^{i\frac{\pi}{4}{\rm sign}\left(\frac{H}{P}\right)}
\sqrt{\frac{K}{8}\left|\frac{H}{P}\right|}\prod_{i=1}^{3}
2i\sin\left(\frac{r_{i}n_{i}+c_{0}}{p_{i}}\right)\right.&
\nonumber\\
&\left.+\sum_{j=0}^{\infty}\frac{1}{j!}(8\pi iK)^{-j}
\left(\frac{P}{H}\right)^{j}
\left.\partial_{\epsilon}^{(2j)}\left[\frac{\prod_{i=1}^{3}
2i\sin\left(2\pi\frac{r_{i}n_{i}+c_{0}+\epsilon}
{p_{i}}\right)}{2i\sin 2\pi \epsilon}
-\frac{\prod_{i=1}^{3}
2i\sin\left(2\pi\frac{r_{i}n_{i}+c_{0}}
{p_{i}}\right)}{4\pi i\epsilon}
\right]\right|_{\epsilon=0}\right\}.&
\label{S.4}
\end{eqnarray}

The contribution of the trivial connection is
\begin{equation}
-\frac{e^{i\frac{\pi}{2}{\rm sign}\left(\frac{H}{P}\right)}}
{\sqrt{8K|H|}}{\rm sign}(P)e^{-\frac{i\pi}{2K}\phi}
\sum_{j=0}^{\infty}\frac{1}{j!}
\left.\left(\frac{\pi}{2iK}\frac{P}{H}\right)^{j}
\left[\partial_{\epsilon}^{(2j)}\frac{\prod_{i=1}^{3}
2i\sin\frac{\epsilon}{p_{i}}}{2i\sin\epsilon}\right]
\right|_{\epsilon=0}.
\label{S.5}
\end{equation}
A remarkable feature of this formula is that its full 2-loop term is
proportional to Casson's invariant~(\ref{4.1.14}).

The formulas analogous to eqs.~(\ref{S.3}), (\ref{S.4}) and
(\ref{S.5}) for the $n$-fibered Seifert manifold are
eqs.~(\ref{5.3.4}), (\ref{5.3.7}) and (\ref{5.3.10}).

\section{Calculation of Witten's Invariant}
\label{*2}
\subsection{Loop Expansion}
\label{*2.1}

We start with a brief description of a stationary phase approximation
to the path integral~(\ref{2}). The stationary phase points are the
extrema of the action (\ref{1}). Since
\begin{equation}
\frac{\delta S}{\delta
A_{\mu}}=\frac{1}{2\pi}\epsilon^{\mu\nu\rho}F_{\nu\rho},
\label{2.1}
\end{equation}
these extrema are flat connections, i.e. connections with
$F_{\mu\nu}=0$. The gauge equivalence classes of flat connections are
in one-to-one correspondence with the homomorphisms
\begin{equation}
\pi_{1}({\cal M})\stackrel{A}{\rightarrow}G,\;\;A:\;x\mapsto g(x)\in G
\label{2.2}
\end{equation}
(G is a gauge group) up to a conjugacy, that is, the homomorphisms
$g(x)$ and $h^{-1}g(x)h$ are considered equivalent.

Each stationary phase point $A^{(i)}$ contributes a classical
exponential $\exp[ikS_{i}]$ times an asymptotic series in $k$:
\begin{equation}
Z({\cal M},k)=
\sum_{i}e^{ikS_{i}}\left(\sum_{n=0}^{\infty}k^{-n}\Delta_{n}^{(i)}\right).
\label{2.3}
\end{equation}
Another form of presenting the same expansion is
\begin{equation}
Z({\cal
M},k)=\sum_{i}\Delta_{0}^{(i)}\exp\;ik\left[S_{i}+\sum_{n=2}^{\infty}
k^{-n}S_{n}^{(i)}\right].
\label{2.4}
\end{equation}
Here $S_{i}$ are Chern-Simons invariants of the flat connections
$A_{\mu}^{(i)}$, and $S_{n}^{(i)}$ are the
$n$-loop quantum corrections
coming from the $n$-loop 1-particle irreducible Feynman diagrams. A
set of Feynman rules for their calculation has been developed in
\cite{AS}, however the actual calculations have been carried out only
up to the 1-loop order.

Generally in quantum field theory a 1-loop factor is an inverse
square root of a determinant  of the second order variations of the
classical action taken at the stationary phase point. However a gauge
invariance of the action (\ref{1}) requires a gauge fixing and an
introduction of the Faddeev-Popov ghost determinant (see \cite{W} for
details). So for the Chern-Simons theory
\begin{equation}
\Delta_{0}=\frac{|\det(k\Delta)|}{\left[\det(-ikL_{-})\right]^{1/2}}.
\label{2.01}
\end{equation}
Here $\Delta$ is a covariant Laplacian
\begin{equation}
\Delta=D_{\mu}D^{\mu},\;\;D_{\mu}=\partial_{\mu}+A_{\mu}
\label{2.02}
\end{equation}
acting on the Lie algebra valued functions, while $L_{-}$ is
\begin{equation}
L_{-}=\left[
\begin{array}{cc}
*d_{A}&-d_{A}*\\
d_{A}*&0
\end{array}\right]
\label{2.03}
\end{equation}
acting on the Lie algebra valued 1-forms and 3-forms. A differential
$d_{A}$ is built upon a covariant derivative $D_{\mu}$, $d_{A}^{2}=0$
for flat connections.

According to \cite{W}, the absolute value of the ratio (\ref{2.01}) is
a square root of the Reidemeister-Ray-Singer torsion $\tau_{R}(A)$. A
detailed expression for the phase of that ratio has been worked out in
\cite{FG}. The 1-loop formula for $Z({\cal M},k)$ presented there is a
sum over the flat connections $A^{(i)}$:
\begin{equation}
Z({\cal M},k)=e^{-i\frac{\pi}{4}(\dim\,G)(1+b^{1})}
\sum_{i}e^{2\pi iKS_{CS}^{(i)}}\tau_{R}^{1/2}e^{-i\frac{\pi}{2}I_{i}}
C_{0}C_{1},
\label{2.5}
\end{equation}
here $K=k+c_{v}$, $c_{v}$ is a dual Coxeter number or,
equivalently, a quadratic Casimir invariant of the adjoint
representation, $b^{1}$ is the first Betti number and $I_{i}$ is a
spectral flow. The factors $C_{0}$ and $C_{1}$ reflect the presence of
the 0-form (i.e. 3-form) and 1-form zero modes in the operators
$\Delta$ and $L_{-}$ of eq.(\ref{2.01}). These factors have to
be slightly modified from their original values in \cite{FG}.

The zero modes are related to the elements of the cohomology spaces
$H^{0}({\cal M},d_{A^{(i)}})$ and $H^{1}({\cal M},d_{A^{(i)}})$. For
each element of $H^{0}$ there is a zero mode of $\Delta$ and a zero
mode of $L_{-}$. For each element of $H^{1}$ there is another zero
mode of $L_{-}$. It is also known that $H^{0}$ can be identified with
a tangent space of the symmetry group $H_{i}$ of the connection
$A^{(i)}_{\mu}$. The group $H_{i}$ consists of the gauge
transformations that do not change $A^{(i)}_{\mu}$. Equivalently,
$H_{i}$ is a subgroup of $G$ whose elements commute with the image of
the homomorphism (\ref{2.2}). As for $H^{1}$, its elements represent
infinitesimal deformations of the connection $A_{\mu}^{(i)}$ which do
not violate the flatness condition. This picture is reminiscent of the
string theory. There the zero modes of the ghosts $c$ and $b$ were
identified with the elements of tangent spaces of the symmetry group
and moduli of the complex structure. However in our case generally not
all the elements of $H^{1}$ can be extended to finite deformations
of flat connections. In other words, $\dim\,H^{1}\leq\dim\,X_{i}({\cal
M})$, where $X_{i}$ is a connected component of the moduli space of
flat connections.

Let us first assume that $\dim\,H^{1}=\dim\,X_{i}$. If operators
$\Delta,L_{-}$ have zero modes, then the Reidemeister torsion can
still be obtained from eq.~(\ref{2.01}) if the zero modes and zero
eigenvalues are removed from there. L. Jeffrey noted in \cite{J} that
$\tau_{R}^{1/2}$ thus defined is an element of
$\Lambda^{\max}H^{0}\otimes(\Lambda^{\max}H^{1})^{*}$. She suggested
to take a canonical element $v\in(\Lambda^{\max}H^{0})^{*}$ derived
from the basic inner product on $H^{0}$ which is a Lie algebra of
$H_{i}$. A pairing of $v$ and $\tau_{R}^{1/2}$ produces a volume form
on the moduli space $X_{i}$. A sum over the flat connections in
eq.~(\ref{2.5}) then includes a natural integration over $X_{i}$.
However, according to \cite{J}, this procedure does not quite agree
with the leading term in the $1/k$ expansion of $Z(S^{3},k)$.

We propose a slightly different prescription. We take any element
$v\in(\Lambda^{\max}H^{0})^{*}$ and balance the integral over $X_{i}$,
defined by pairing of $v$ and $\tau_{R}^{1/2}$, with a factor of
$1/{\rm Vol}\,(H_{i})$, volume of $H_{i}$ being defined by the same element
$v$\footnote{
The factor $1/{\rm Vol}\,(H_{i})$ appeared in slightly different
circumstances in \cite{W2}. It also appeared in \cite{RS1} and
\cite{RS2} where the Alexander polynomial was produced from a
Chern-Simons theory based on a supergroup $U(1|1)$. It was shown there
that ${\rm Vol}\,(U(1|1))=0$, so that the flat connections for which
$H_{i}=U(1|1)$, gave infinite contributions to the partition function.
These infinities helped to explain the nonmultiplicativity of the
Alexander polynomial which distinguished it from the family of the
$SU(N)$ Jones polynomials.}.
We show in Appendix why this factor should appear after the removal of
the zero modes from eq.~(\ref{2.01}) by considering a simple finite
dimensional version of a gauge invariant path integral. We also
demonstrate in the end of subsection \ref{*2.3} how our prescription
fits the value of $Z(S^{3},k)$.

There is another consequence of dropping the zero modes from the
determinants in eq.~(\ref{2.01}). Each nonzero mode of the operator
$\Delta$ carries a factor of $k$ in eq.~(\ref{2.01}) and each nonzero
mode of $L_{-}$ carries there a factor\footnote{Actually
a partition function~(\ref{2}) has also a factor
$k^{\#\,{\rm of\,all\,modes\,of}\,\Delta
-\frac{1}{2}\#\,{\rm of\,all\,modes\,of}\,L_{-}}$ hidden in the
integration measure ${\cal D}A_{\mu}$.}
of $(-ik)^{-1/2}$. By dropping
the modes, we loose these factors. Therefore dropping an element of
$H^{0}$ produces an extra factor $(ik)^{-1/2}$ while dropping an
element of $H^{1}$ creates a factor $(-ik)^{1/2}$. Thus
\begin{equation}
C_{0}=\frac{1}{{\rm Vol}\,(H_{i})}(ik)^{-(\dim\,H^{0})/2}.
\label{2.6}
\end{equation}
We could also assume that
\begin{equation}
C_{1}=(-ik)^{(\dim\,H^{1})/2}
\label{2.7}
\end{equation}
However the 1-form zero modes of $L_{-}$ that can not be extended to
finite deformations of the flat connection, should not be simply
dropped from eq.~(\ref{2.01}). A nonzero mode of $L_{-}$ contributes
to eq.~(\ref{2.01}) through a gaussian integral
\begin{equation}
\int_{-\infty}^{+\infty}\exp(i\pi k\lambda x^{2})dx\sim (-ik)^{-1/2}
\label{2.8}
\end{equation}
A zero mode 1-form  that hits obstruction contributes through the
integral
\begin{equation}
\int_{-\infty}^{+\infty}\exp(i\pi k\lambda x^{4})dx\sim(-ik)^{-1/4}
\label{2.9}
\end{equation}
Therefore a corrected version of the formula for $C_{1}$ is
\begin{equation}
C_{1}=(-ik)^{\frac{\dim\,H^{1}}{2}}
(-ik)^{-\frac{\dim\,H^{1}-\dim\,X_{i}}{4}}=
(-ik)^{\frac{\dim\,H^{1}+\dim\,X_{i}}{4}}
\label{2.10}
\end{equation}
and the 1-loop formula~(\ref{2.5}) takes the form
\begin{eqnarray}
Z({\cal M},k)&=&
\sum_{i}\exp\left(2\pi iKS_{CS}^{(i)}\right)
\exp-i\frac{\pi}{4}\left[(1+b^{1})\dim\,G+2I_{i}+\dim\,H^{0}+
\frac{\dim\,H^{1}+\dim\,X_{i}}{2}\right]
\nonumber\\
&&\times
\frac{1}{{\rm Vol}\;(H_{i})}\tau_{R}^{1/2}k^{-\frac{\dim\,H^{0}}{2}+
\frac{\dim\,H^{1}+\dim\,X_{i}}{4}}.
\label{2.005}
\end{eqnarray}
%

%*************************************
\subsection{Surgery Calculus}
\label{*2.2}
%*************************************

Here we briefly present Witten's recipe of an exact calculation of the
partition function (\ref{2}). Witten used the fact that the Hilbert
space of the Chern-Simons quantum field theory is isomorphic to the
space of conformal blocks of the level $k$ 2-dimensional WZW model
based on the same group $G$. More specifically, a Chern-Simons
Hilbert space corresponding to a 2-dimensional torus is equivalent to
the space of affine characters of $G$ (see, e.g. \cite{EMSS}).

Consider a path integral (\ref{2}) calculated over a solid torus with
a Wilson line carrying representation
$V_{\Lambda}$ of $G$ going inside it. $\Lambda$ denotes the shifted
highest weight, i.e. the highest weight of $V_{\Lambda}$ is
$\Lambda-\rho$, $\rho$ being half the sum of positive roots of $G$:
$\rho= \frac{1}{2}\sum_{\lambda_{i}\in\Delta_{+}}\lambda_{i}$.
An inclusion of the Wilson line means that the integrand of
eq.~(\ref{2}) is multiplied by a trace of a holonomy
${\rm tr}_{V_{\Lambda}}{\rm Pexp}\left(\oint
A_{\mu}dx^{\mu}\right)$. Such integral is a function of the boundary
conditions imposed on $A_{\mu}$ on the boundary of the solid torus.
Therefore it is an element $|\Lambda\rangle$ of the Hilbert space of
$T^{2}$. Witten claimed that this element corresponds to the affine
character of level $k$ built upon $V_{\Lambda}$ and that all such
elements corresponding to the integrable affine representations form
an orthonormal basis in that Hilbert space.

The group $SL(2,{\bf Z})$ acting as modular transformations on
$T^{2}$, generates canonical transformations in the phase space of the
classical Chern-Simons theory. Therefore $SL(2,{\bf Z})$ can be
unitarily represented in the Hilbert space. This representation is
determined by the action of the matrices $S$ and $T$
\begin{equation}
S=\left(
\begin{array}{cc}
0&-1\\
1&0\end{array}\right),\;\;
T=\left(
\begin{array}{cc}
1&1\\0&1\end{array}\right)
\label{2.2.1}
\end{equation}
on the affine characters.
An action of a general unimodular matrix
\begin{equation}
M^{(p,q)}=\left(\begin{array}{cc}p&r\\q&s\end{array}\right)
\in SL(2,{\bf Z})
\label{2.2.2}
\end{equation}
is determined by its presentation as a product
\begin{equation}
M^{(p,q)}=T^{a_{t}}S\ldots T^{a_{1}}S.
\label{2.2.3}
\end{equation}
The integer numbers $a_{i}$ form a continued fraction expansion of
$p/q$:
\begin{equation}
\frac{p}{q}=a_{t}-\frac{1}{a_{t-1}-\frac{1}{\ldots-\frac{1}{a_{1}}}}.
\label{2.2.4}
\end{equation}
For more details on this construction see e.g. \cite{FG} and \cite{J}.

A lens space $L(p,q)$ can be constructed by gluing the boundaries of
two solid tori. The boundaries are identified trivially after a
matrix $SM^{(p,-q)}$ acts on one of them, i.e. that matrix determines
how the boundaries are glued together. Before the gluing, each solid
torus produces an affine character $V_{\rho}$ growing out of the
trivial representation, as a state on its boundary. Hence
according to the postulates of a quantum field theory, a partition
function (\ref{2}) is equal to a matrix element
\begin{equation}
Z(L(p,q),k)=(\tilde{S}\tilde{M}^{(p,-q)})_{\rho\rho}
\label{2.2.5}
\end{equation}
here tilde denotes a representation of the $SL(2,{\bf Z})$ matrices in
the space of affine characters.

Note that the lens space depends only
on the numbers $p$ and $q$. Different choices of the entries $r$ and
$s$ of the matrix (\ref{2.2.2}) correspond to different framings of
the same lens space. A framing is a choice of three vector fields
which form a basis in the tangent space at each point of the manifold.
A phase of a partition function $Z$ depends on a choice of
framing. Formula (\ref{2.005}) gives a 1-loop approximation to
$Z({\cal M},k)$ in the standard framing. The surgical formulas should
also be reduced to the standard framing in order to
yield a true invariant of the manifold. We will discuss this reduction
in the end of this subsection and in subsection~\ref{*3.3}.

Consider now a manifold $S^{2}\times S^{1}$. A Seifert manifold
$X(\frac{p_{1}}{q_{1}},\ldots,\frac{p_{n}}{q_{n}})$ is constructed by
cutting out
the tubular neighborhoods of $n$ strands going parallel to
$S^{1}$ and then gluing them back after performing the
$M^{(p_{i},q_{i})}$ transformations on their boundaries\footnote{It is
not hard to see that $X(\frac{p}{q})$ and
$X(\frac{p_{1}}{q_{1}},\frac{p_{2}}{q_{2}})$ are the lens spaces.}.
These transformations change the states on the surfaces of the solid
tori from $|\rho\rangle$ into
\begin{equation}
|\Lambda^{\prime}_{i}\rangle=\sum_{\Lambda}|\Lambda\rangle
M^{(p_{i},q_{i})}_{\Lambda \rho}
\label{2.2.6}
\end{equation}
Therefore an invariant of a Seifert manifold is given by a multiple
sum
\begin{equation}
Z(X,k)=\sum_{\Lambda_{1},\ldots,\Lambda_{n}}
M^{(p_{1},q_{1})}_{\Lambda_{1}
\rho}\ldots M^{(p_{n},q_{n})}_{\Lambda_{n}\rho}
N_{\Lambda_{1}\ldots\Lambda_{n}}
\label{2.2.7}
\end{equation}
Here $N_{\Lambda_{1}\ldots\Lambda_{n}}$ is a Verlinde number, which is
equal to the invariant of the manifold $S^{2}\times S^{1}$
containing $n$ Wilson lines carrying representations
$V_{\Lambda_{i}}$ along $S^{1}$. This number is also ``almost''
equal to the number of times that a trivial representation appears in
a decomposition of a tensor product
$\bigotimes_{i=1}^{n}V_{\Lambda_{i}}$.
$N_{\Lambda_{1}\ldots\Lambda_{n}}$ will be equal to that number if
$V_{\Lambda_{i}}$ are representations of the quantum algebra
$G_{q}$. An expression for $N_{\Lambda_{1}\ldots\Lambda_{n}}$ in the
case of $n=3$ and $G=SU(2)$ is presented in subsection \ref{*3.1}, a
general case will be considered in subsection \ref{*5.1}.

As in the case of a lens space, the phase of a partition function
(\ref{2.2.7}) also depends on the choice of numbers $r_{i},s_{i}$.
Witten found in \cite{W} that a change of framing by one unit is
accompanied by a change in that phase by $\frac{\pi c}{12}$, $c$ being
a central charge of the level $k$ WZW model.

To get a partition function $Z(X,k)$ in the standard framing, the
r.h.s. of eq.~(\ref{2.2.7}) should be multiplied by a phase factor
determined in \cite{FG} to be equal to
\begin{equation}
\exp\frac{i\pi c}{12}\left[-3\sigma+\sum_{i=1}^{n}\sum_{j=1}^{t_{i}}
a_{j}^{(i)}\right].
\label{2.2.8}
\end{equation}
Here $a_{j}^{(i)}$ form a continued fraction expansion of
$p_{i}/q_{i}$, while
\begin{equation}
\sigma=-{\rm sign}\left(\sum_{i=1}^{n}\frac{q_{i}}{p_{i}}\right)
+\sum_{i=1}^{n}{\rm sign}\left(\frac{p_{i}}{q_{i}}\right)+
\sum_{i=1}^{n}\sum_{j=1}^{t_{i}-1}{\rm sign}(a_{j}^{(i)})
\label{2.2.9}
\end{equation}
%

%*****************************************************
\subsection{$SU(2)$ Formulas and Poisson Resummation}
\label{*2.3}
%*****************************************************
Explicit formulas for the $SL(2,{\bf Z})$ representation in the space
of affine $SU(2)$ characters of level $k$ were derived in \cite{J}.
There are $k+1$ integrable $\widetilde{SU}(2)$ representations with
spins $0\leq j\leq\frac{k}{2}$.
We use the shifted highest weight $\alpha=2j+1$ instead of
$j$ and $K=k+2$ instead of $K$, so that $0<\alpha<K$. The weight
$\rho$ is equal to 1 for $SU(2)$.

The formulas for $\tilde{S}_{\alpha\beta}$ and
$\tilde{T}_{\alpha\beta}$ are well known:
\begin{equation}
\tilde{S}_{\alpha\beta}=\sqrt{\frac{2}{K}}\sin\frac{\pi\alpha\beta}{K},\;\;
\tilde{T}_{\alpha\beta}=e^{-\frac{i\pi}{4}}
e^{\frac{i\pi}{2K}\alpha^{2}}\delta_{\alpha\beta}
\label{2.3.1}
\end{equation}
A substitution of these expressions in the r.h.s. of eq.~(\ref{2.2.3})
turns it into a multiple finite gaussian sum. A summation over the
intermediate indices goes from 1 to $K-1$. An application of
the Poisson
resummation formula in \cite{J} converted that sum into another
gaussian sum with a summation interval independent of $K$:
\begin{equation}
\tilde{M}^{(p,q)}_{\alpha\beta}= i\frac{{\rm sign}(q)}{\sqrt{2K|q|}}
e^{-\frac{i\pi}{4}\Phi(M^{(p,q)})}\sum_{\mu=\pm
1}\sum_{n=0}^{q-1}\mu\exp\frac{i\pi}{2Kq}\left[p\alpha^{2}-
2\mu\alpha(\beta+2Kn)+s(\beta+2Kn)^{2}\right]
\label{2.3.2}
\end{equation}
Here $\Phi(M)$ is a Rademacher phi function defined as follows
\begin{equation}
\Phi\left[\begin{array}{cc}p&r\\q&s\end{array}\right]=\left\{
\begin{array}{lll}
\frac{p+s}{q}-12s(s,q)&\;{\rm if}&\;q\neq0\\
\frac{r}{s}&\;{\rm if}&\;q=0\end{array}\right.,
\label{2.3.3}
\end{equation}
a function $s(s,q)$ being a Dedekind sum:
\begin{equation}
s(m,n)=\frac{1}{4n}\sum_{j=1}^{n-1}\cot\frac{\pi j}{n}\cot\frac{\pi
mj}{n}.
\label{2.3.4}
\end{equation}

We illustrate the use of the Poisson formula
\begin{equation}
\sum_{n\in {\bf Z}}f(n)=\sum_{m\in{\bf Z}}\int_{-\infty}^{+\infty}
e^{2\pi imx}f(x)dx
\label{2.3.5}
\end{equation}
by explicitly deriving an expression for the matrix element
\begin{equation}
(\tilde{S}\tilde{T}^{p}\tilde{S})_{\alpha\beta}=-
\frac{e^{-\frac{i\pi}{4}p}}{2K}\sum_{\gamma=1}^{K-1}\sum_{\mu_{1},\mu_{2}=\pm1}
\mu_{1}\mu_{2}\exp\frac{i\pi}{2K}\left[p\gamma^{2}+2\gamma(\mu_{1}\alpha+
\mu_{2}\beta)\right]
\label{2.3.6}
\end{equation}
This excercise will prepare us for the calculations that we will
perform in section~\ref{*3}.

We have to extend the summation range of $\gamma$ to ${\bf Z}$ in
order to be able  to use eq.~(\ref{2.3.5}). The summand in
eq.~(\ref{2.3.6}) is even and periodic with a period of $2K$. We first
double the range of
summation:
$\sum_{\gamma=1}^{K-1}\rightarrow\frac{1}{2}\sum_{\gamma=-K}^{K-1}$.
Then a formula
\begin{equation}
\sum_{n=0}^{N-1}f(n)=N\lim_{\epsilon\rightarrow 0}\epsilon^{1/2}
\sum_{n\in{\bf Z}}e^{-\pi\epsilon n^{2}}f(n),\;\;{\rm if}\;
f(n)=f(n+N),
\label{2.3.7}
\end{equation}
and a Poisson resummation  allow us to transform eq.~(\ref{2.3.6})
into
\begin{eqnarray}
(\tilde{S}\tilde{T}^{p}\tilde{S})_{\alpha\beta}=-
\frac{e^{-\frac{i\pi}{4}p}}{2K}\lim_{\epsilon\rightarrow 0}
(K\epsilon^{1/2})\sum_{\gamma\in{\bf Z}}
\sum_{\mu_{1,2}=\pm 1}\mu_{1}\mu_{2}e^{-\pi\epsilon\gamma^{2}}
\exp\frac{i\pi}{2K}\left[p\gamma^{2}+2\gamma
(\mu_{1}\alpha+\mu_{2}\beta)\right]\nonumber\\
=-
\frac{e^{-\frac{i\pi}{4}p}}{2K}\lim_{\epsilon\rightarrow 0}
(K\epsilon^{1/2})\sum_{n\in{\bf Z}}\int_{-\infty}^{+\infty}d\gamma
\sum_{\mu_{1,2}=\pm 1}\mu_{1}\mu_{2}e^{-\pi\epsilon\gamma^{2}}
\exp\frac{i\pi}{2K}\left[p\gamma^{2}+2\gamma
(\mu_{1}\alpha+\mu_{2}\beta+2Kn)\right]
\label{2.3.8}
\end{eqnarray}
Note that a change from a sum to an integral over $\gamma$ has been
essentially accomplished through a substitution
\begin{equation}
\beta\rightarrow\beta+2Kn
\label{2.3.9}
\end{equation}
and a subsequent summation over $n$. This is a trick that works for a
general expression (\ref{2.2.3}). Just one substitution like
(\ref{2.3.9}) for an initial or final index converts all the
intermediate sums in eq.~(\ref{2.2.3}) into gaussian integrals.
This is, in fact, the origin of the expression $(\beta+2Kn)$ in
eq.~(\ref{2.3.2}).

{}From this point we can proceed in two ways.
A straightforward way is to integrate
the r.h.s. of eq.~(\ref{2.3.8}) over $\gamma$. Then, after neglecting
some irrelevant terms we get a formula
\begin{equation}
(\tilde{S}\tilde{T}^{p}\tilde{S})_{\alpha\beta}=-e^{-\frac{i\pi}{4}p}
\sqrt{\frac{iK}{2p}}\lim_{\epsilon\rightarrow\ 0}\sum_{n\in{\bf Z}}
\sum_{\mu_{1,2}=\pm1}\mu_{1}\mu_{2}\epsilon^{1/2}
e^{-4\pi\epsilon\frac{K^{2}}{p^{2}}n^{2}}
\exp\left[-\frac{i\pi}{2Kp}(\mu_{1}\alpha+\mu_{2}\beta+2Kn)^{2}\right]
\label{2.3.10}
\end{equation}
The second exponential here is periodic in $n$ with a period $p$.
Therefore a second application of eq.~(\ref{2.3.7}), this time
backwards, leads to the final expression
\begin{equation}
(\tilde{S}\tilde{T}^{p}\tilde{S})_{\alpha\beta}=-e^{-\frac{i\pi}{4}p}
\sqrt{\frac{i}{8Kp}}\sum_{n=0}^{p-1}
\sum_{\mu_{1,2}=\pm1}\mu_{1}\mu_{2}
\exp\left[-\frac{i\pi}{2Kp}(\mu_{1}\alpha+\mu_{2}\beta+2Kn)^{2}\right]
\label{2.3.11}
\end{equation}

An equivalent way to treat the r.h.s. of eq.~(\ref{2.3.8}) is to
notice
that since an integral over $\gamma$ is gaussian, a stationary phase
approximation is exact. An array of stationary phase points
$\gamma_{\rm st}(n)=-2Kn/p$ and their contributions exhibit the same
symmetry properties as a summand of eq.~(\ref{2.3.6}). Therefore an
inverse use of eq.~(\ref{2.3.7}) shows that we can drop a factor
$(K\epsilon^{1/2})$ from the r.h.s. of eq.~(\ref{2.3.8}) and restrict
the sum there to those values of $n$ for which
\begin{equation}
0\leq \gamma_{\rm st}(n)\leq K
\label{2.3.110}
\end{equation}
The formula that we get this way is slightly different in its form
from eq.~(\ref{2.3.11}). For $p$ -- odd  we get
\begin{eqnarray}
(\tilde{S}\tilde{T}^{p}\tilde{S})_{\alpha\beta}
&=&-e^{-\frac{i\pi}{4}p}
\sqrt{\frac{i}{2Kp}}
\sum_{\mu_{1,2}=\pm1}\mu_{1}\mu_{2}
\left[\frac{1}{2}
\exp\left(-\frac{i\pi}{2Kp}(\mu_{1}\alpha+\mu_{2}\beta)^{2}\right)
\right.\nonumber\\
&&\left.+\sum_{n=1}^{\frac{p-1}{2}}
\exp\left(-\frac{i\pi}{2Kp}(\mu_{1}\alpha+\mu_{2}\beta+2Kn)^{2}\right)
\right]
\label{2.3.111}
\end{eqnarray}
The $1/2$ factor in front of the first exponential here is due to the
fact that the stationary point $\gamma_{\rm st}(0)=0$ is on the
boundary of the interval (\ref{2.3.110}). There is another such
stationary point $\gamma_{\rm st}(p/2)$ for $p$ -- even.

In the next section we will apply eqs.~(\ref{2.3.5}) and (\ref{2.3.7})
to formula~(\ref{2.2.7}). Meanwhile we use eq.~(\ref{2.3.11})
together with eq.~(\ref{2.2.5}) in order to get an expression for
$Z(L(p,-1),k)$ and check a factor of $1/{\rm Vol}(H)$ in the 1-loop
approximation formula~(\ref{2.005}). In the large $k$ limit for
odd~$p$
\begin{equation}
Z(L(p,-1),k)=(\tilde{S}\tilde{T}^{p}\tilde{S})_{11}\approx
e^{-\frac{i\pi}{4}p}\left[\sqrt{2}\pi\left(\frac{i}{Kp}\right)^{3/2}
+4\sqrt{\frac{i}{2Kp}}\sum_{n=1}^{\frac{p-1}{2}}
e^{-2\pi iKn^{2}/p}\sin^{2}\frac{2\pi n}{p}\right]
\label{2.3.13}
\end{equation}
The first term in the square brackets is a contribution of the trivial
connection, while the remaining sum goes over nontrivial maps $\pi_{1}
(L(p,-1))=Z_{p}\rightarrow SU(2)$. The $SU(2)$ subgroup $H$ commuting
with the image of $\pi_{1}$ is $SU(2)$ and $U(1)$ respectively.

According to \cite{FG}, a square root of the Reidemeister torsion of
the trivial connection is $p^{-3/2}$ and that of a nontrivial one is
$4p^{-1}\sin^{2}(2\pi n/p)$. Therefore if eq.~(\ref{2.005}) is
correct, then
\begin{equation}
{\rm
Vol}\left(SU(2)\right)=\frac{1}{\pi\sqrt{2}},\;\;{\rm
Vol}\left(U(1)\right)=\sqrt{2}.
\label{2.3.14}
\end{equation}
The same value of ${\rm Vol}\left(SU(2)\right)$ is predicted by a
large $k$ limit of $Z(S^{3},k)$, which is equal to $\sqrt{2}\pi
k^{-3/2}$. The group $SU(2)$ is a 3-dimensional sphere and $U(1)$ is
its big circle. We get both volumes~(\ref{2.3.14}) if we assume that
the radius of that sphere is $1/(\sqrt{2}\pi)$.
This value is not unnatural. The path integral measure in quantum
theories contains an implicit factor $(2\pi\hbar)^{-1/2}$ coming with
each of the 1-dimensional integrals comprising the path integral. For
the Chern-Simons theory $\hbar=\pi/k$. The radius of $SU(2)$ was
calculated in the ``reduced'' measure, so its actual value is $1$.

%*****************************************************
\section{A Large $k$ Limit of the Invariants of 3-Fibered Seifert
Manifolds}
\label{*3}
%*****************************************************

%*****************************************************
\subsection{Stationary Phase Points}
\label{*3.1}
%*****************************************************
Let us try to apply the Poisson formula~(\ref{2.3.5}) to the partiton
function~(\ref{2.2.7}) for the case of $n=3$ in order to put it in the
form~(\ref{2.3}) or (\ref{2.4}).

We start by giving an explicit expression for
$N_{\alpha_{1},\alpha_{2},\alpha_{3}}$ inside the fundamental cube
\begin{equation}
0<\alpha_{1},\alpha_{2},\alpha_{3}<K
\label{3.1}
\end{equation}
According to its definition, $N_{\alpha_{1},\alpha_{2},\alpha_{3}}=1$
iff $\alpha_{1}+\alpha_{2}+\alpha_{3}$ is odd and the following 4
inequalities are satisfied:
\begin{eqnarray}
\alpha_{1}+\alpha_{2}-\alpha_{3}&>&0\nonumber\\
\alpha_{1}-\alpha_{2}+\alpha_{3}&>&0\nonumber\\
-\alpha_{1}+\alpha_{2}+\alpha_{3}&>&0\nonumber\\
\alpha_{1}+\alpha_{2}+\alpha_{3}&<&2K
\label{3.3}
\end{eqnarray}
Otherwise, $N_{\alpha_{1},\alpha_{2},\alpha_{3}}=0$. We can drop a
restriction on the parity of $\alpha_{1}+\alpha_{2}+\alpha_{3}$ if we
change the formula~(\ref{2.2.7}) into
\begin{equation}
Z\left(X(\frac{p_{1}}{q_{1}},\frac{p_{2}}{q_{2}},
\frac{p_{3}}{q_{3}}),k\right)=
\sum_{0<\alpha_{1},\alpha_{2},\alpha_{3}<K}\sum_{\lambda=0,\frac{1}{2}}
\frac{1}{2}e^{2\pi i\lambda(1-\alpha_{1}-\alpha_{2}-\alpha_{3})}
\tilde{M}^{(p_{1},q_{1})}_{\alpha_{1}1}
\tilde{M}^{(p_{2},q_{2})}_{\alpha_{2}1}
\tilde{M}^{(p_{3},q_{3})}_{\alpha_{3}1}
\tilde{N}_{\alpha_{1},\alpha_{2},\alpha_{3}}
\label{3.4}
\end{equation}
here $\tilde{N}_{\alpha_{1},\alpha_{2},\alpha_{3}}$  is a modified
Verlinde number, $\tilde{N}_{\alpha_{1},\alpha_{2},\alpha_{3}}=1$  in
the whole region defined by the inequalities~(\ref{3.3}). This region
is a tetrahedron within the fundamental cube~(\ref{3.1}) (see~Fig.~1).
To simplify this picture we draw its section by a plane
$\alpha_{3}={\rm const}$ in Fig.~2 (region~$1_{+}$).

At this stage we could use a discrete stationary phase approximation
method described in Appendix A of \cite{FG}, in order to get the
1-loop approximation of $Z$. The stationary phase points inside the
tetrahedron would correspond to the irreducible flat connections. The
conditional stationary phase points on the faces of the tetrahedron
would correspond to the reducible flat connections.

We use a different appoach close to the one in subsection \ref{*2.3}
in order to get the full $1/k$ expansion of $Z(X,k)$. We want to
extend the sum in eq.~(\ref{3.4}) from the fundamental cube to the
whole 3-dimensional space. We do this in two steps by using the
symmetries of the matrices $\tilde{M}^{(p,q)}_{\alpha,\beta}$ under
the affine Weyl transformations. These transformations include the
ordinary Weyl reflections as well as the shifts by the root lattice
multiplied by $K$:
\begin{equation}
\tilde{M}^{(p,q)}_{-\alpha,\beta}=-\tilde{M}^{(p,q)}_{\alpha,\beta},\;\;
\tilde{M}^{(p,q)}_{\alpha+2K,\beta}=\tilde{M}^{(p,q)}_{\alpha,\beta}
\label{3.5}
\end{equation}
The easiest way to see these symmetries is to use eq.~(\ref{2.2.3})
and expressions~(\ref{2.3.1}) for $\tilde{S}_{\alpha\beta}$ and
$\tilde{T}_{\alpha\beta}$.

The first of eqs.~(\ref{3.5}) enables us to extend the sum~(\ref{3.4})
to a bigger cube:
$\sum_{0<\alpha_{1},\alpha_{2},\alpha_{3}<K}\rightarrow
\frac{1}{8}\sum_{-K<\alpha_{1},\alpha_{2},\alpha_{3}<K}$ if we extend
$\tilde{N}_{\alpha_{1},\alpha_{2},\alpha_{3}}$ as an antisymmetric
function inside that cube (see the regions $2_{-},3_{+}$ and $4_{-}$
in Fig.~2). The translational invariance of $\tilde{M}_{\alpha\beta}$
together with eq.~(\ref{2.3.7}) brings us to another formula for $Z$:
\begin{equation}
Z=\lim_{\epsilon\rightarrow 0}
\frac{(2K\epsilon^{1/2})^{3}}{8}
\sum_{\alpha_{1},\alpha_{2},\alpha_{3}\in{\bf Z}}
\sum_{\lambda=0,\frac{1}{2}} \frac{1}{2}e^{2\pi
i\lambda(1-\alpha_{1}-\alpha_{2}-\alpha_{3})
-\pi\epsilon(\alpha_{1}^{2}+\alpha_{2}^{2}+\alpha_{3}^{2})}
\tilde{M}^{(p_{1},q_{1})}_{\alpha_{1}1}
\tilde{M}^{(p_{2},q_{2})}_{\alpha_{2}1}
\tilde{M}^{(p_{3},q_{3})}_{\alpha_{3}1}
\tilde{N}_{\alpha_{1},\alpha_{2},\alpha_{3}}
\label{3.7}
\end{equation}
if we require $\tilde{N}_{\alpha_{1},\alpha_{2},\alpha_{3}}$ to be a
periodic function of its indices with the period of $2K$ (see Fig.~3).

A Poisson formula~(\ref{2.3.5}) transforms the sum over $\alpha_{i}$
into an integral:
\begin{equation}
\sum_{\alpha_{1},\alpha_{2},\alpha_{3}\in{\bf Z}}=
\int d\alpha_{1}d\alpha_{2}d\alpha_{3}
\sum_{n_{1},n_{2},n_{3}\in{\bf Z}}
\prod_{i=1}^{3}\delta(\alpha_{i}-n_{i})=
\int d\alpha_{1}d\alpha_{2}d\alpha_{3}
\sum_{m_{1},m_{2},m_{3}\in{\bf Z}}
\exp\left(2\pi i\sum_{i=1}^{3}\alpha_{i}m_{i}\right)
\label{3.8}
\end{equation}
On the other hand,
\begin{equation}
e^{2\pi i\alpha m}\tilde{M}_{\alpha\beta}=
\tilde{M}_{\alpha,\beta+2Kqm},
\label{3.9}
\end{equation}
so the sum over $m_{i}$ and the exponential in the r.h.s. of
eq.~(\ref{3.8}) can be absorbed by extending the sum over $n$ in
eq.~(\ref{2.3.2}) to all integer numbers. Finally
\begin{eqnarray}
Z(X,k)&=&Z_{1}Z_{2},
\label{3.10}\\
Z_{1}&=&i^{3}{\rm sign}(q_{1}q_{2}q_{3})
\exp\left[-\frac{i\pi}{4}\sum_{i=1}^{3}\Phi(M^{(p_{i},q_{i})})\right],
\label{3.11}\\
Z_{2}&=&\lim_{\epsilon\rightarrow 0}\frac{(2K\epsilon^{1/2})^{3}}{8}
\frac{1}{2}\sum_{\lambda=1,\frac{1}{2}}e^{2\pi i\lambda}
\int \tilde{N}_{\alpha_{1}\alpha_{2}\alpha_{3}}
\prod_{i=1}^{3}\frac{d\alpha_{i}}{\sqrt{2K|q_{i}|}}
\sum_{\mu_{i}=\pm 1}\sum_{n_{i}\in{\bf Z}}
\mu_{i}e^{-\pi\epsilon\alpha_{i}^{2}}
\nonumber\\
&&\times\exp\frac{i\pi}{2Kq_{i}}\left[
p_{i}\alpha_{i}^{2}-2\alpha_{i}(2K(n_{i}+q_{i}\lambda)+\mu_{i})
+s_{i}(2Kn_{i}+\mu_{i})^{2}\right].
\label{3.12}
\end{eqnarray}

The integral in eq.~(\ref{3.12}) is gaussian, but the function
$\tilde{N}_{\alpha_{1}\alpha_{2}\alpha_{3}}$ carves a rather
complicated region out of the 3-dimensional $\alpha$-space. A slice
of that region for $\alpha_{3}={\rm const}$ is depicted in Fig. 3.
Fortunately,
this region can be represented as a linear combination of
positive strips (double wedges in 3-dimensional space)
\begin{equation}
\alpha_{1}-\alpha_{3}+2Kl<\alpha_{2}<\alpha_{1}+\alpha_{3}+2Kl,\;\;
l\in{\bf Z}
\label{3.13}
\end{equation}
and negative strips
\begin{equation}
\alpha_{3}-\alpha_{1}+2Kl<\alpha_{2}<-\alpha_{3}-\alpha_{1}+2Kl,\;\;
l\in{\bf Z}
\label{3.14}
\end{equation}
Each strip (double wedge) is in turn a difference between two
half-planes (half-spaces). Overall we have a superposition of
half-spaces
\begin{equation}
\sum_{i=1}^{3}\nu_{i}\alpha_{i}+2Kl>0,
\;\;\nu_{1}=-1,\;\nu_{2},\nu_{3}=\pm 1
\label{3.15}
\end{equation}
These half-spaces are related to the Bernstein-Gelfand -Gelfand
resolution of affine modules. We will use this relation in subsection
\ref{*5.1}.

A sign of the contribution coming from a half-space~(\ref{3.15}) is
determined by the product $\nu_{2}\nu_{3}$. Therefore we can change
$\int \tilde{N}_{\alpha_{1}\alpha_{2}\alpha_{3}}$ in eq.~(\ref{3.12})
for
\begin{equation}
-\sum_{n\in{\bf Z}}\sum_{\nu_{2,3}=\pm 1}\nu_{1}\nu_{2}\nu_{3}
\int_{\sum_{i=1}^{3}\nu_{i}\alpha_{i}+2Kl>0}
\label{3.16}
\end{equation}

The stationary points of the phase in eq.~(\ref{3.12}) are
\begin{equation}
\alpha_{i}^{({\rm st})}=2K\frac{\tilde{n}_{i}}{p_{i}},\;\;
{\rm here}\;\;\tilde{n}_{i}=n_{i}+q_{i}\lambda
\label{3.17}
\end{equation}
There are also conditional stationary points on the boundary planes
\begin{equation}
\sum_{i=1}^{3}\nu_{i}\alpha_{i}+2Kl=0
\label{3.171}
\end{equation}
They are
\begin{equation}
\alpha_{i}^{({\rm cst})}=\frac{2K}{p_{i}}
\nu_{i}(n_{i}-q_{i}c_{0}),
\label{3.18}
\end{equation}
here
\begin{equation}
c_{0}=\frac{H}{P}\sum_{j=1}^{3}\frac{n_{j}}{p_{j}}
+l,\;
P=p_{1}p_{2}p_{3},\;H=P\sum_{j=1}^{3}\frac{q_{i}}{p_{i}}=
p_{1}p_{2}q_{3}+p_{1}q_{2}p_{3}+q_{1}p_{2}p_{3}.
\label{3.19}
\end{equation}
$|H|$ is the order of homology group of the Seifert manifold
$X\left(\frac{p_{1}}{q_{1}},\frac{p_{2}}{q_{2}},\frac{p_{3}}{q_{3}}\right)$.
The points~(\ref{3.18}) form a 2-dimensional lattice on the
plane~(\ref{3.171}). Note that $\alpha_{i}^{({\rm cst})}$ are not
changed under a simultaneous shift
\begin{equation}
n_{i}\longrightarrow n_{i}+q_{i}m,\;\;m\in{\bf Z}
\label{3.210}
\end{equation}

Consider now an integral from eq.~(\ref{3.12}) with a
substitution~(\ref{3.16}):
\begin{equation}
\int_{\sum_{i=1}^{3}\nu_{i}\alpha_{i}+2Kl>0}
\prod_{i=1}^{3}\frac{d\alpha_{i}}{\sqrt{2K|q_{i}|}}
\exp\frac{i\pi}{2Kq_{i}}\left[p_{i}\alpha_{i}^{2}-
2\alpha_{i}(2K(n_{i}+q_{i}\lambda)+\mu_{i})+
s_{i}(2Kn_{i}+\mu_{i})^{2}\right]
\label{3.20}
\end{equation}
We dropped a regularization factor
$exp(-\pi\epsilon\sum_{i=1}^{3}\alpha_{i}^{2})$, while keeping in mind
that it will suppress a contribution of the stationary
points~(\ref{3.17}) and~(\ref{3.18}) by its value at those points. If
a point~(\ref{3.17}) does not belong to the half-space
of~(\ref{3.20}), then the integral is equal to a contribution of the
conditional point~(\ref{3.18}). If, however, a point~(\ref{3.17}) is
within the half-space, then we use an obvious relation
\begin{equation}
\int_{\sum_{i=1}^{3}\nu_{i}\alpha_{i}+2Kl>0}
=\int_{{\bf R}^{3}}-
\int_{\sum_{i=1}^{3}\nu_{i}\alpha_{i}+2Kl<0}
\label{3.21}
\end{equation}
The first integral in the r.h.s. of this equation is purely gaussian,
it is determined by the point~(\ref{3.17}). The second integral is
again determined by a conditional point~(\ref{3.18}). We will
calculate both integrals in the next subsection. Here we just note
that as it follows from eq.~(\ref{3.21}), a contribution of a
point~(\ref{3.17}) to the integral~(\ref{3.20}) is either zero or a
quantity which does not depend on the half-space to which it belongs.
Therefore if a point~(\ref{3.17}) belongs to an array of tetrahedra
whose slice is depicted in Fig. 3, then its contribution to the whole
expression~(\ref{3.12}) is equal to the first integral in the r.h.s.
of eq.~(\ref{3.21}). If the point does not belong to the array, then
its contribution is zero.

The overall picture is this: we have two lattices~(\ref{3.17}) and
(\ref{3.18}). $Z_{2}$ is equal to the sum of the contributions of the
points of these lattices. The lattices and the contributions of their
points exhibit the same symmetry under the affine Weyl group
transformations, as the summand of eq.~(\ref{3.7}). Therefore by using
the inverse eq.~(\ref{2.3.7}) in exactly the same way as we did in
deriving eq.~(\ref{2.3.111}), we drop the factor
$(2K\epsilon^{1/2})^{3}/8$ from the integral in eq.~(\ref{3.12}). At
the same time we restrict the sum over $n_{i}$ to those stationary
points~(\ref{3.17}) which belong to the fundamental
tetrahedron~(\ref{3.3}) and to those conditional stationary
points~(\ref{3.18}) which lie on its faces.

The stationary points that belong to the intersection of the
planes~(\ref{3.171}) require a special care. Their contribution to the
integral over the region carved by
$\tilde{N}_{\alpha_{1}\alpha_{2}\alpha_{3}}$ is proportional to the
number of planes to which they belong. However the reduction to the
fundamental tetrahedron should also account for the fact that these
points are invariant under the action of a subgroup $W^{\prime}$ of
the affine Weyl group. Therefore the total contribution of such points
is equal to their contribution to the  integral~(\ref{3.12}) times a
factor
\begin{equation}
\frac{\#\;{\rm of\; planes}}{\#\;{\rm of\; elements\; in}\;W^{\prime}}
\label{3.220}
\end{equation}
This factor is similar to the factor $1/2$ in eq.~(\ref{2.3.111}). It
is equal to $\frac{2}{2}=1$ for the points on the edges of the
tetrahedron and to $\frac{4}{8}=\frac{1}{2}$ for the points on the
vertices of the tetrahedron.

We can ``unfold'' the surface of the tetrahedron and require the
points~(\ref{3.18}) to belong to the intersection of the plane
%
%\begin{equation}
$
\alpha_{1}+\alpha_{2}+\alpha_{3}=0
%\label{3.22}
%\end{equation}
$
with the cube $-2K<\alpha_{1}<0,\;-2K<\alpha_{2}<0,\;0<\alpha_{3}<2K$.
This intersection is an equilateral triangle (see Fig. 4) consisting
of 4 smaller triangles that can be mapped by Weyl reflections onto the
faces of the fundamental tetrahedron.

There is yet another way to view the fundamental set of conditional
stationary phase points. As we have noted, the triplets $n_{i}$
related by a transformation~(\ref{3.210}) define the same point
through
eqs.~(\ref{3.18}). A transformation
\begin{equation}
n_{i}\longrightarrow n_{i}+mp_{i},\;\;
n_{j}\longrightarrow n_{j}-mp_{j},\;\;
i\neq j
\label{3.23}
\end{equation}
does not change $c_{0}$ and shifts $\alpha_{i}^{({\rm cst})}$ by $2Km$
and $\alpha_{j}^{({\rm cst})}$ by $-2Km$, thus leaving them within the
same equivalence class of affine Weyl transformations. We can describe
a fundamental region of the conditional stationary phase points as a
factor of a lattice of all integer triplets $n_{i}$ over a lattice
generated by three vectors
\begin{equation}
\vec{v}_{1}=(q_{1},q_{2},q_{3}),\;
\vec{v}_{2}=(p_{1},-p_{2},0),\;
\vec{v}_{3}=(0,p_{2},-p_{3})
\label{3.24}
\end{equation}
The number of triplets $n_{i}$ inside that factor is equal to the
volume of a parallelepiped formed by the vectors $\vec{v}_{i}$
\begin{equation}
\#\;{\rm of\;conditional\;points}=\left|
\begin{array}{ccc}
q_{1}&q_{2}&q_{3}\\
p_{1}&-p_{2}&0\\
0&p_{2}&-p_{3}\end{array}\right|=|H|
\label{3.25}
\end{equation}
We should be interested only in approximately half of the triplets, because
the volume of the prism built upon a triangle of Fig.~4 is twice as small as
that of the parallelepiped which is built upon the whole parallelogram.
Thus the number of the conditional stationary phase points within the
fundamental domain is approximately equal to half the rank of the homology
group. This result is not surprising since we intend to identify the
conditional stationary phase points with reducible $SU(2)$ flat
connections. The number of these connections is also approximately equal
to $|H|/2$.

%*****************************************************
\subsection{The Integrals}
\label{*3.2}
%*****************************************************
\noindent
\underline{Stationary Phase Points}
\nopagebreak

We start with the simplest case of a contribution of the
point~(\ref{3.17}) which is inside the fundamental
tetrahedron~(\ref{3.3}). As we saw in the previous subsection, it is
equal to the gaussian integral taken over the whole $\alpha$-space:
\begin{eqnarray}
Z_{2}&=&\frac{1}{2}\sum_{\lambda=0,\frac{1}{2}}e^{2\pi i\lambda}
\prod_{i=1}^{3}\sum_{\mu_{i}=\pm 1}\mu_{i}\int_{-\infty}^{+\infty}
\frac{d\alpha_{i}}{\sqrt{2K|q_{i}|}}
\exp\frac{i\pi}{2Kq_{i}}\left[p_{i}\alpha_{i}^{2}-
2\alpha_{i}(2K\tilde{n}_{i}+\mu_{i})+
s_{i}(2Kn_{i}+\mu_{i})^{2}\right]\nonumber\\
&=&Z_{3}\frac{1}{2}\sum_{\lambda=0,\frac{1}{2}}e^{2\pi i\lambda}
\prod_{i=1}^{3}
\frac{1}{\sqrt{|p_{i}|}}2i\sin2\pi\left(\frac{r_{i}}{p_{i}}
\tilde{n}_{i}+s_{i}\lambda\right)
\exp2\pi
iK\left[\frac{r_{i}}{p_{i}}\tilde{n}_{i}^{2}-q_{i}s_{i}\lambda^{2}
\right].
\label{3.2.1}
\end{eqnarray}
Here $Z_{3}$ is a factor that will be present in all the subsequent
expressions for the contributions to $Z_{2}$:
\begin{equation}
Z_{3}=\prod_{i=1}^{3}e^{\frac{i\pi}{4}{\rm sign}(p_{i}q_{i})}
\exp\left(\frac{i\pi}{2K}\frac{r_{i}}{p_{i}}\right)
\label{3.2.01}
\end{equation}
\noindent
\underline{Conditional Stationary Phase Points}
\nopagebreak

Consider now a contribution of the points
\begin{equation}
\alpha^{({\rm cst})}_{i}=\frac{2K}{p_{i}}(n_{i}-q_{i}c_{0}),\;\;
c_{0}=\frac{P}{H}\sum_{i=1}^{3}\frac{n_{i}}{p_{i}}
\label{3.2.2}
\end{equation}
which belong to the plane
\begin{equation}
\alpha_{1}+\alpha_{2}+\alpha_{3}=0
\label{3.2.3}
\end{equation}
We start with an integral~(\ref{3.2.1}) that we take over the region
\begin{equation}
\alpha_{1}+\alpha_{2}+\alpha_{3}<0
\label{3.2.4}
\end{equation}
We can return to the integral over the whole $\alpha$-space if we
introduce an extra factor
\begin{equation}
\theta(-\alpha_{1}-\alpha_{2}-\alpha_{3})=
\int_{0}^{+\infty}dx\int_{-\infty}^{+\infty}dc\exp\left[2\pi ic(
\alpha_{1}+\alpha_{2}+\alpha_{3}+x)\right]
\label{3.2.5}
\end{equation}
The invariance of $\alpha^{({\rm cst})}_{i}$ under the
transformation~(\ref{3.210}) implies that we should make a
substitution~(\ref{3.210}) in the integral~(\ref{3.2.1}) and take a
sum over all integer $m$. As a result, the full contribution of the
point  $\alpha^{({\rm cst})}_{i}$ to $Z_{2}$ is
\begin{eqnarray}
Z_{2}&=&\frac{1}{2}\sum_{\lambda=0,\frac{1}{2}}e^{2\pi i\lambda}
\sum_{m\in{\bf Z}}\int_{0}^{+\infty}
dx\int_{-\infty}^{+\infty}dc\;
e^{2\pi icx}\prod_{i=1}^{3}\sum_{\mu_{i}=\pm 1}\mu_{i}
\int_{-\infty}^{+\infty}\frac{d\alpha_{i}}
{\sqrt{2K|q|}}
\nonumber\\
&&\times
\exp\frac{i\pi}{2Kq_{i}}\left[p_{i}\alpha_{i}^{2}-
2\alpha_{i}\left(2K(n_{i}+q_{i}(\lambda+m-c))+\mu_{i}\right)
+s_{i}\left(2K(n_{i}+q_{i}m)+\mu_{i}\right)^{2}\right]
\nonumber\\
&=&Z_{3}\frac{e^{-\frac{i\pi}{4}{\rm sign}(H/P)}}
{\sqrt{2K|H|}}\exp2\pi iK\left[
\sum_{i=1}^{3}\frac{r_{i}}{p_{i}}n_{i}^{2}+\frac{H}{P}c_{0}^{2}\right]
\sum_{\mu_{1,2,3}=\pm 1}\left\{\prod_{i=1}^{3}\mu_{i}
\exp\left[2\pi i\mu_{i}\frac{r_{i}n_{i}+c_{0}}{p_{i}}\right]\right\}
\nonumber\\
&&\times
\frac{1}{2}\sum_{\lambda=0,\frac{1}{2}}e^{2\pi i\lambda}
\sum_{m\in{\bf Z}}I(m),
\label{3.2.6}
\end{eqnarray}
here
\begin{equation}
I(m)=\int_{0}^{+\infty}dx\exp\left[2\pi ix(c_{0}+m+\lambda)+
\frac{i\pi}{2K}\frac{P}{H}\left(x+\sum_{i=1}^{3}\frac{\mu_{i}}
{p_{i}}\right)^{2}\right]
\label{3.2.7}
\end{equation}
We calculate the integral~(\ref{3.2.7}) in the spirit of remarks
preceding eq.~(\ref{3.21}). If $\frac{P}{H}(c_{0}+m+\lambda)<0$, then
the stationary point of the phase in eq.~(\ref{3.2.7}) lies outside
the integration region, and the dominant contribution comes from the
boundary point $x=0$:
\begin{eqnarray}
I(m)&=&\sum_{j=0}^{\infty}\frac{1}{j!}\left(\frac{i\pi}{2K}
\frac{P}{H}\right)^{j}\int_{0}^{\infty}dx\left(x+
\sum_{i=1}^{3}\frac{\mu_{i}}{p_{i}}\right)^{2j}
\exp[2\pi ix(c_{0}+m+\lambda)]\nonumber\\
&=&\sum_{j=0}^{\infty}\frac{1}{j!}(8\pi iK)^{-j}
\left(\frac{P}{H}\right)^{j}\partial_{\epsilon}^{(2j)}
\left.\left[\int_{0}^{\infty}dx\;\exp\left(
2\pi ix(c_{0}+m+\lambda+\epsilon)+
2\pi i\epsilon\sum_{i=1}^{3}\frac{\mu_{i}}{p_{i}}
\right)\right]\right|_{\epsilon=0}.
\label{3.2.8}
\end{eqnarray}
If however $\frac{P}{H}(c_{0}+m+\lambda)>0$, then we use a relation
similar to eq.~(\ref{3.21}):
\begin{equation}
\int_{0}^{\infty}dx=\int_{-\infty}^{+\infty}dx\;-\;\int_{-\infty}^{0}dx
\label{3.2.9}
\end{equation}
The integral $\int_{-\infty}^{+\infty}dx$ is fully determined by
the stationary phase point and therefore has been accounted for in
eq.~(\ref{3.2.1}). The integral $-\int_{-\infty}^{0}dx$ is dominated
by the boundary point $x=0$ and leads to the same
expression~(\ref{3.2.8}). Thus eq.~(\ref{3.2.8}) is valid if
$c_{0}+m+\lambda\neq 0$.

A Poisson formula~(\ref{2.3.5}) allows us to convert a sum over $m$
into a ``discretization'' of the integral over $x$:
\begin{eqnarray}
\frac{1}{2}\sum_{\lambda=0,\frac{1}{2}}e^{2\pi i\lambda}
\sum_{m\in{\bf Z}}I(m)&=&
\sum_{j=0}^{\infty}\frac{1}{j!}(8\pi iK)^{-j}
\left(\frac{P}{H}\right)^{j}\partial_{\epsilon}^{(2j)}
\left.\left[\frac{1}{2}\sum_{\lambda=0,\frac{1}{2}}
e^{2\pi i\left(\lambda+\epsilon\sum_{i=1}^{3}
\frac{\mu_{i}}{p_{i}}\right)}\sum_{x=0}^{\infty}
e^{2\pi ix(c_{0}+\lambda+\epsilon)}\right]
\right|_{\epsilon=0}
\nonumber\\
&=&-\sum_{j=0}^{\infty}\frac{1}{j!}(8\pi
iK)^{-j}\left(\frac{P}{H}\right)\partial_{\epsilon}^{(2j)}
\left.\left[\frac{e^{2\pi
i\epsilon\sum_{i=1}^{3}\frac{\mu_{i}}{p_{i}}}}
{2i\sin 2\pi(c_{0}+\epsilon)}\right]\right|_{\epsilon=0}
\label{X.1}
\end{eqnarray}
By substituting this expression into eq.~(\ref{3.2.6}) and taking a
sum over $\mu_{i}$ we get the final expression for the contribution of
the stationary point~(\ref{3.2.2}) to $Z_{2}$:
\begin{eqnarray}
Z_{2}&=&-Z_{3}\frac{e^{-i\frac{\pi}{4}{\rm
sign}\left(\frac{H}{P}\right)}}{\sqrt{2K|H|}}
\exp 2\pi iK\left[\sum_{i=1}^{3}\frac{r_{i}}{p_{i}}n_{i}^{2}+
\frac{H}{P}c_{0}^{2}\right]
\nonumber\\
&&\times\sum_{j=0}^{\infty}\frac{1}{j!}(8\pi iK)^{-j}
\left(\frac{P}{H}\right)^{j}\left.\left[\partial_{c}^{(2j)}
\frac{\prod_{i=1}^{3}2i\sin\left(2\pi\frac{r_{i}n_{i}+c}{p_{i}}
\right)}{2i\sin 2\pi c}\right]\right|_{c=c_{0}}.
\label{3.2.12}
\end{eqnarray}

\noindent
\underline{A Stationary Phase Point on the Boundary}
\nopagebreak

A stationary phase point~(\ref{3.17}) presents a special case when it
belongs to the boundary of one of the planes~(\ref{3.15}). Suppose,
that $\alpha_{i}^{({\rm cst})}$ satisfy conditions~(\ref{3.2.3}).
Comparing eqs.~(\ref{3.17}) and (\ref{3.2.2}), we see that this may
happen if
\begin{equation}
%$
c_{0}+\lambda=0
%$
\label{3.2.13}
\end{equation}
or, in other words, $c_{0}=0,-\frac{1}{2}$.

We can proceed with the same analysis as for an ordinary conditional
stationary phase point up to eq.~(\ref{3.2.8}). The integral $I(0)$
requires a separate calculation. According to eq.~(\ref{3.2.7}),
\begin{eqnarray}
I(0)&=&\int_{0}^{\infty}dx\exp\left[\frac{i\pi}{2K}\frac{P}{H}
\left(x+\sum_{i=1}^{3}\frac{\mu_{i}}{p_{i}}\right)^{2}\right]
\nonumber\\
&=&\sqrt{\frac{K}{2}\left|\frac{H}{P}\right|}
e^{i\frac{\pi}{4}{\rm sign}(H/P)}-
\int_{0}^{\sum_{i=1}^{3}\frac{\mu_{i}}{p_{i}}}
dx\;\exp\left[\frac{i\pi}{2K}\frac{P}{H}x^{2}\right]
\nonumber\\
&=&\sqrt{\frac{K}{2}\left|\frac{H}{P}\right|}
e^{i\frac{\pi}{4}{\rm sign}(H/P)}-
\sum_{j=0}^{\infty}\frac{1}{j!}(8\pi iK)^{-j}
\left(\frac{P}{H}\right)^{j}\left.\left[\partial_{\epsilon}^{(2j)}
\frac{e^{2\pi i\sum_{i=1}^{3}\frac{\mu_{i}}{p_{i}}}-1}
{2\pi i\epsilon}\right]\right|_{\epsilon=0}.
\label{X.2}
\end{eqnarray}
The remaining part of eq.~(\ref{X.1}) is equal to
\begin{equation}
-\sum_{j=0}^{\infty}\frac{1}{j!}(8\pi iK)^{-j}
\left(\frac{P}{H}\right)e^{2\pi ic_{0}}\partial_{\epsilon}^{(2j)}
\left.\left[e^{2\pi i\epsilon\sum_{i=1}^{3}\frac{\mu_{i}}{p_{i}}}
\left(\frac{1}{2i\sin 2\pi\epsilon}-\frac{1}{4\pi i\epsilon}
\right)\right]\right|_{\epsilon=0}.
\label{X.3}
\end{equation}
Adding $I(0)$ with an extra factor $\frac{1}{2}e^{2\pi ic_{0}}$ to
this expression and substituting it into eq.~(\ref{3.2.6}) brings us
to the formula
\begin{eqnarray}
Z_{2}&=&Z_{3}\frac{e^{-i\frac{\pi}{4}{\rm
sign}\left(\frac{H}{P}\right)}}
{\sqrt{2K|H|}}\exp 2\pi iK\left[\sum_{i=1}^{3}
\frac{r_{i}}{p_{i}}n_{i}^{2}+\frac{H}{P}c_{0}^{2}\right]
e^{2\pi ic_{0}}
\nonumber\\
&&\times\left\{e^{i\frac{\pi}{4}{\rm sign}\left(\frac{H}{P}\right)}
\sqrt{\frac{K}{8}\left|\frac{H}{P}\right|}
\prod_{i=1}^{3}2i\sin\left(2\pi\frac{r_{i}n_{i}+c_{0}}{p_{i}}
\right)\right.
\label{3.2.16}
%\nonumber
\\
&&-\left.\sum_{j=0}^{\infty}\frac{1}{j!}(8\pi iK)^{-j}
\left(\frac{P}{H}\right)^{j}\partial_{\epsilon}^{(2j)}
\left.\left[\frac{\prod_{i=1}^{3}2i\sin\left(2\pi
\frac{r_{i}n_{i}+c_{0}+\epsilon}{p_{i}}\right)}
{2i\sin(2\pi\epsilon)}-\frac{\prod_{i=1}^{3}2i\sin
\left(2\pi\frac{r_{i}n_{i}+c_{0}}{p_{i}}\right)}{4\pi i\epsilon}
\right]\right|_{\epsilon=0}\right\}.
\nonumber
%\label{3.2.16}
\end{eqnarray}

\noindent
\underline{Trivial Connection}
\nopagebreak

In the case of $n_{1}=n_{2}=n_{3}=c_{0}=0$
the formula~(\ref{3.2.16}) requires an extra $1/2$ factor coming from
the ratio~(\ref{3.220}). After some simplifications it becomes
\begin{equation}
Z_{2}=-Z_{3}\frac{e^{-i\frac{\pi}{4}{\rm
sign}\left(\frac{H}{P}\right)}}{\sqrt{8K|H|}}
\sum_{j=0}^{\infty}\frac{1}{j!}\left(\frac{\pi}{2iK}
\frac{P}{H}\right)^{j}\left.\left[\partial_{\epsilon}^{(2j)}
\frac{\prod_{i=1}^{3}2i\sin\left(\frac{\epsilon}{p_{i}}\right)}
{2i\sin\epsilon}\right]\right|_{\epsilon=0}.
\label{3.2.17}
\end{equation}
%

%*****************************************************
\subsection{Framing Corrections}
\label{*3.3}
%*****************************************************

We have to reduce our formulas for $Z$ to the standard framing in
order to obtain a true topological invariant that is independent, for
example, of the choice of $r_{i}$ and $s_{i}$ in the matrices
$M^{(p_{i},q_{i})}$. We add an extra phase factor~(\ref{2.2.8})  to
$Z$. According to \cite{J},
\begin{equation}
\sum_{j=1}^{t_{i}}a_{j}^{(i)}-3\sum_{j=1}^{t_{i}-1}{\rm
sign}\left(a_{j}^{(i)}\right)
=\Phi(M^{(p_{i},q_{i})}).
\label{3.3.1}
\end{equation}
Also note that
\begin{equation}
\Phi(M^{(p_{i},q_{i})})-3{\rm sign}(p_{i}q_{i})=
\Phi(SM^{(p_{i},q_{i})})
\label{3.3.2}
\end{equation}
The central charge of the $SU(2)$ WZW model is
\begin{equation}
c=\frac{3(K-2)}{K}.
\label{3.3.3}
\end{equation}
Therefore the full framing correction is
\begin{eqnarray}
Z_{f}&=&\exp\frac{i\pi}{4}\left[
\sum_{i=1}^{3}\left(\Phi(M^{(p_{i},q_{i})})-
3{\rm sign}(p_{i}q_{i})\right)
+3{\rm sign}\left(\frac{H}{P}\right)\right]
\nonumber\\
&&\times\exp-\frac{i\pi}{2K}\left[\sum_{i=1}^{3}
\Phi(SM^{(p_{i},q_{i})})+
3{\rm sign}\left(\frac{H}{P}\right)\right]
\label{3.3.4}
\end{eqnarray}
so that a corrected version of the product of phase
factors $Z_{1}Z_{2}$ is
\begin{equation}
Z_{1}Z_{3}Z_{f}=
e^{i\frac{3\pi}{4}{\rm sign}\left(\frac{H}{P}\right)}
\prod_{i=1}^{3}{\rm sign}(p_{i})
\exp-\frac{i\pi}{2K}\left[3{\rm sign}\left(\frac{H}{P}\right)+
\sum_{i=1}^{3}\left(12s(q_{i},p_{i})-\frac{q_{i}}{p_{i}}\right)\right].
\label{3.3.5}
\end{equation}
We used the following property of the Dedekind sum in order to derive
this formula:
\begin{equation}
s(m^{*},n)=s(m,n),\;{\rm if}\;mm^{*}=1({\rm mod}\;n).
\label{3.3.6}
\end{equation}
Eq.~(\ref{3.3.5}) together with eqs.~(\ref{3.2.01}), (\ref{3.2.12}),
(\ref{3.2.16}) and (\ref{3.2.17}) leads to the final
formulas~(\ref{S.1})--(\ref{S.5}).

%*****************************************************
\section{One-Loop Approximation Formulas}
\label{*4}
%*****************************************************

%*****************************************************
\subsection{Irreducible Flat Connections}
\label{*4.1}
%*****************************************************

Irreducible flat connections on a Seifert manifold
$X\left(\frac{p_{1}}{q_{1}},\ldots,\frac{p_{n}}{q_{n}}\right)$
are the ones for which the subgroup $H$ commuting with the image of
the homomorphism~(\ref{2.2}) does not have continuous parameters. In
the case of $G=SU(2)$ this simply means that the image of~(\ref{2.2})
is noncommutative.

The fundamental group $\pi_{1}$ of the Seifert manifold is generated
by the elements $x_{1},\ldots,x_{n},h$ satisfying relations
\begin{equation}
x_{i}^{p_{i}}h^{q_{i}}=1,\;hx_{i}=x_{i}h,\;\prod_{i=1}^{n}x_{i}=1.
\label{4.1.1}
\end{equation}
The elements $x_{i}$ go
around the solid tori that make up the manifold, while $h$ goes along
the $S^{1}$ cycle of the ``mother-manifold'' $S^{1}\times S^{2}$
(see \cite{FS} and \cite{FG} for details).

Suppose that the image of $h$ does not belong to the center of
$SU(2)$. Since $h$ commutes with all the elements of $\pi_{1}$, then
the whole image of $\pi_{1}$ belongs to $U(1)\subset SU(2)$. Therefore
$H\supset U(1)$, so this is a reducible case. An irreducible
connection is produced only if the image of $h$ belongs to the center
of $SU(2)$:
\begin{equation}
h\stackrel{A}{\mapsto}e^{2\pi i\lambda}
\left(\begin{array}{cc}
1&0\\0&-1\end{array}\right),\;\;
\lambda=0,\frac{1}{2}.
\label{4.1.2}
\end{equation}
The images of the elements $x_{i}$ belong to the conjugation classes
of diagonal matrices whose phases we denote as $u_{i}$:
\begin{equation}
x_{i}\stackrel{A}{\mapsto}g_{i}^{-1}
\left(
\begin{array}{cc}
e^{2\pi iu_{i}}&0\\0&e^{-2\pi iu_{i}}
\end{array}
\right)g_{i}
\label{4.1.3}
\end{equation}
The first of relations~(\ref{4.1.1}) determines the possible values of
these phases:
\begin{equation}
u_{i}=\frac{\tilde{n}_{i}}{p_{i}},
\label{4.1.4}
\end{equation}
here the numbers $\tilde{n}_{i}$ are defined in eq.~(\ref{3.17}). The
phases $u_{i}$ determine the map~(\ref{2.2}) uniquely up to an overall
conjugation if the number of the solid tori is $n=3$. If $n>3$, then
each particular choice of the phases $u_{i}$ corresponds to a
connected component of the $2(n-3)$ dimensional moduli space of
these maps, which is a moduli space of flat connections on a Seifert
manifold. Such connected component is isomorphic to the space of flat
connections on an $n$-holed sphere if the holonomies around the holes
are fixed by~eq.~(\ref{4.1.3}). We will discuss this subject further
in subsection~\ref{*5.2}. Here we specialize to the case $n=3$ and
present the expressions for the manifold invariants entering
eq.~(\ref{2.005}).
A Chern-Simons invariant of an irreducible flat connection on a
Seifert manifold was computed in~\cite{A}:
\begin{equation}
S_{CS}=\sum_{i=1}^{3}\frac{1}{p_{i}}
(r_{i}n_{i}^{2}-2\lambda n_{i}-q_{i}\lambda^{2})
=\sum_{i=1}^{3}\left(\frac{r_{i}}{p_{i}}\tilde{n}_{i}^{2}-
q_{i}s_{i}\lambda^{2}\right)\;({\rm mod}\;1).
\label{4.1.5}
\end{equation}
According to \cite{F}, a square root of the corresponding Reidemeister
torsion is
\begin{equation}
\tau_{X}^{1/2}=\prod_{i=1}^{3}\frac{2}{\sqrt{p_{i}}}|\sin(2\pi\phi_{i})|,
\label{4.1.6}
\end{equation}
here
\begin{equation}
\phi_{i}=\frac{r_{i}n_{i}-\lambda}{p_{i}}=
\frac{r_{i}}{p_{i}}\tilde{n}_{i}+s_{i}\lambda\;({\rm mod}\;1)
\label{4.1.7}
\end{equation}
are the phases of the conjugation classes of the holonomies along the
central fibers of the solid tori that make up the Seifert manifold.

The spectral flow was calculated in~\cite{FS}:
\begin{equation}
I_{A}=-3+8S_{CS}+\sum_{i=1}^{3}\frac{2}{p_{i}}
\sum_{l=1}^{p_{i}-1}\cot\left(\frac{\pi r_{i}l}{p_{i}}\right)
\cot\left(\frac{\pi l}{p_{i}}\right)
\sin^{2}\left[\frac{2\pi l}{p_{i}}(r_{i}n_{i}-\lambda)\right].
\label{4.1.8}
\end{equation}
L. Jeffrey presented in her paper~\cite{J} a proof by D. Zagier of an
equation
\begin{equation}
(-i){\rm sign}\left(\sin\frac{2\pi rn}{p}\sin\frac{2\pi n}{p}\right)
=\exp\frac{i\pi}{2}\left[
\frac{8rn^{2}}{p}-\frac{2}{p}\sum_{l=1}^{p-1}
\cot\frac{\pi l}{p}\cot\frac{\pi ql}{p}
\sin^{2}\frac{2\pi nl}{p}\right],
\label{4.1.9}
\end{equation}
here $n,p,q,r\in{\bf Z},\;qr=-1\;({\rm mod}\;p)$. A slight
modification of that proof shows that eq.~(\ref{4.1.9}) works also for
half-integer $n$ if we multiply its l.h.s. by an extra factor of
$e^{2\pi in}$. Then an application of eq.~(\ref{4.1.9}) with a
substitution
\begin{equation}
q=r_{i},\;\;r=q_{i},\;\;n=r_{i}n_{i}-\lambda
\label{4.1.10}
\end{equation}
to the r.h.s. of eq.~(\ref{4.1.8}) leads to a formula for the
exponential of the spectral flow:
\begin{equation}
\exp\left(-\frac{i\pi}{2}I_{A}\right)=
-\prod_{i=1}^{3}e^{2\pi i\lambda}{\rm sign}\left(
\sin\frac{2\pi\tilde{n}_{i}}{p_{i}}\sin2\pi\phi_{i}\right)=
-e^{2\pi i\lambda}\prod_{i=1}^{3}{\rm
sign}\left(\sin 2\pi\phi_{i}\right),
\label{4.1.11}
\end{equation}
here we used the fact that
$0<\frac{\tilde{n}_{i}}{p_{i}}<\frac{1}{2}$ because
$0<\alpha^{({\rm st})}_{i}<K$. Apparently the role of the factor
$(-i)^{I_{A}}$ is to remove the absolute value from the sines in the
square root of the Reidemeister torsion~(\ref{4.1.6}). A similar
effect was observed for lens spaces in~\cite{J}.

For an irreducible connection on a 3-fibered Seifert manifold
$X\left(\frac{p_{1}}{q_{1}},\frac{p_{2}}{q_{2}},\frac{p_{3}}{q_{3}}\right)$,
$\dim H^{0}=\dim H^{1}=b^{1}=0$ and $\dim SU(2)=3$, while $H$ is a
center of $SU(2)$ which consists of 2 elements, ${\rm Vol}(H)=2$.
Therefore according to eq.~(\ref{2.005}), the 1-loop contribution of an
irreducible flat connection should be equal to
\begin{equation}
-\frac{1}{2}e^{-i\frac{3\pi}{4}}e^{2\pi i \lambda}
\prod_{i=1}^{3}\frac{2}{\sqrt{|p_{i}|}}
\sin 2\pi\left(\frac{r_{i}}{p_{i}}\tilde{n}_{i}+s_{i}\lambda\right)
\exp2\pi iK\left(\frac{r_{i}}{p_{i}}\tilde{n}_{i}^{2}
-q_{i}s_{i}\lambda^{2}\right).
\label{4.1.12}
\end{equation}
If we compare this expression to eq.~(\ref{S.1}), then we see
that\footnote{in assumption of $p_{1},p_{2},p_{3},H>0$} the exact
contribution differs from eq.~(\ref{4.1.12}) only by a phase factor
\begin{equation}
\exp-\frac{i\pi}{2K}\left[3{\rm sign}\left(\frac{H}{P}\right)+
\sum_{i=1}^{3}\left(12s(q_{i},p_{i})-\frac{q_{i}}{p_{i}}\right)\right].
\label{4.1.13}
\end{equation}
It comes from the overall phase factor $Z_{1}Z_{3}Z_{f}$ and can be
interpreted as a 2-loop correction according to eq.~(\ref{2.4}). Note
that this factor is the same for all the stationary phase
contributions. It does not ``feel'' the background gauge field and
seems to be of ``gravitational'' origin. A similar 2-loop phase factor
has been found in~\cite{G} and~\cite{J} for the lens spaces $L(p,q)$
to be equal to $\exp\left[\frac{i\pi}{2K}12s(q,p)\right]$. S.
Garoufalidis noted in~\cite{G}, that this phase is proportional to
Casson's invariant extended by K. Walker to rational homology spheres.
According to \cite{Wa}, this invariant is equal to $s(q,p)$ for the
lens space $L(p,q)$.
C. Lescop computed the Casson-Walker invariant for $n$-fibered Seifert
manifolds in~\cite{L}:
\begin{equation}
\lambda_{CW}\left(X(\frac{p_{1}}{q_{1}},\ldots,\frac{p_{n}}{q_{n}})\right)
=\frac{1}{12}\frac{P}{H}\left(2-n+\sum_{i=1}^{n}p_{i}^{-2}\right)
-\frac{1}{12}\left[3{\rm sign}\left(\frac{H}{P}\right)+
\sum_{i=1}^{n}\left(12s(q_{i},p_{i})-\frac{q_{i}}{p_{i}}\right)\right].
\label{4.1.14}
\end{equation}
We see that the phase of~(\ref{4.1.13}) is indeed proportional to the
second term in the r.h.s. of eq.~(\ref{4.1.14}), however the first
term (which is dominating in the limit of large $p_{i}$) is missing.
We will see that the missing part appears in the total 2-loop
correction to the trivial connection, which includes some terms of the
asymptotic series together with the phase~(\ref{4.1.13}).

%************************************************
\subsection{General Reducible Flat Connections}
\label{*4.2}
%************************************************

As we noted in the previous subsection,
the image of $\pi_{1}$ under the
homomorphism~(\ref{2.2}) belongs to the $U(1)$ subgroup of $SU(2)$ for
the reducible flat connections. This generally happens when the image
of $h$ does not belong to the center of $SU(2)$:
\begin{equation}
h\stackrel{A}{\mapsto}
\left(\begin{array}{cc}e^{2\pi ic_{0}}&0\\
0&e^{-2\pi ic_{0}}\end{array}\right)
,\;\;c_{0}\neq 0,\frac{1}{2}.
\label{4.2.1}
\end{equation}
Since $h$ commutes with all $x_{i}$, their images should also be
diagonal. The first of equations~(\ref{4.1.1}) again determines the
phases:
\begin{equation}
x_{i}\stackrel{A}{\mapsto}
\left(\begin{array}{cc}e^{2\pi iu_{i}}&0\\
0&e^{-2\pi iu_{i}}\end{array}\right)
,\;\;u_{i}=\frac{n_{i}-q_{i}c_{0}}{p_{i}},
\label{4.2.2}
\end{equation}
while $c_{0}$ is determined by the second equation in~(\ref{3.2.2}).
The phases of the holonomies going along the fibers of the solid tori
are
\begin{equation}
\phi_{i}=\frac{r_{i}n_{i}+c_{0}}{p_{i}}
\label{4.2.3}
\end{equation}

A Chern-Simons action and a square root of the Reidemester torsion are
known to be\footnote{see e.g. \cite{RS2} where these quantities were
calculated by using a $U(1|1)$ Chern-Simons-Witten theory}
\begin{equation}
S_{CS}=\sum_{i=1}^{3}\frac{r_{i}}{p_{i}}n_{i}^{2}+\frac{H}{P}c_{0}^{2}
,\;\;
\tau_{X}^{1/2}=|H|^{-1/2}
\left|\frac{\prod_{i=1}^{3}2\sin(2\pi\phi_{i})}{2\sin(2\pi c_{0})}
\right|.
\label{4.2.5}
\end{equation}
Reducibility of connection means that this time
$\dim H^{0}=1,\;H=U(1),\;{\rm Vol}(H)=1/\sqrt{2}$.

All these formulas are compatible with the leading term in $1/k$
expansion of the conditional
stationary phase contribution~(\ref{S.3})
at least up to a phase factor. Indeed,
we see that the 1-loop part of eq.(\ref{S.3})
(assuming that $p_{1},p_{2},p_{3},H>0$)
is equal to
\begin{equation}
\frac{i}{\sqrt{2KH}}\frac{
\prod_{i=1}^{3}2\sin(2\pi\phi_{i})}
{2\sin(2\pi c_{0})}
\exp2\pi iK\left[\sum_{i=1}^{3}\frac{r_{i}}{p_{i}}n_{i}^{2}+
\frac{H}{P}c_{0}\right]
\label{4.2.6}
\end{equation}
%

%************************************************
\subsection{Special Reducible Connections}
\label{*4.3}
%************************************************

The special reducible flat connections are those for which one or more
sines in eq.~(\ref{4.2.5}) are equal to zero. This amounts to a
condition that a stationary phase point $\alpha^{({\rm st})}$ defined
by eq.~(\ref{3.17}) belongs to a face, an edge or a vertex of the
fundamental tetrahedron~(\ref{3.3}).

\noindent
\underline{Point on a Face}
\nopagebreak

Suppose that a condition~(\ref{3.2.13}) is satisfied for some value of
$\lambda$. This means that the element $h\in\pi_{1}$ is mapped ,
according to eq.~(\ref{4.2.1}), to $e^{2\pi i\lambda}
\left(\begin{array}{cc}1&0\\0&1\end{array}\right)$, so that the
conditional stationary phase point on a face of the tetrahedron is, in
fact, unconditional. The approximation~(\ref{4.2.6}) breaks down, the
reason being that eq.~(\ref{S.4}) should be used instead of
eq.~(\ref{S.3}). Then the leading contribution to a partition
function is equal to one-half of eq.~(\ref{4.1.12}):
\begin{equation}
\frac{1}{4}e^{-i\frac{3\pi}{4}}e^{2\pi i \lambda}
\prod_{i=1}^{3}\frac{2}{\sqrt{|p_{i}|}}
\sin 2\pi\left(\frac{r_{i}}{p_{i}}\tilde{n}_{i}+s_{i}\lambda\right)
\exp2\pi iK\left(\frac{r_{i}}{p_{i}}\tilde{n}_{i}^{2}
-q_{i}s_{i}\lambda^{2}\right).
\label{4.3.1}
\end{equation}
Let us reconcile this expression with eq.~(\ref{2.005}). The fact that a
denominator of eq.~(\ref{4.2.5}) is zero for $c_{0}=0,\frac{1}{2}$
indicates a presence of 1-form zero modes in the operator $L_{-}$.
Indeed, the first two conditions~(\ref{4.1.1}) fix the images of
$x_{i}$ in $SU(2)$ only up to arbitrary conjugations because the
image of $h$ again belongs to the center of $SU(2)$. The last
condition~(\ref{4.1.1}) says that the points $x_{1},x_{2}x_{1}$ and
$x_{3}x_{2}x_{1}=1$ form a ``curved'' triangle inside $SU(2)$. The
size of the sides of this triangle is fixed by the first
condition~(\ref{4.1.1}), but their orientation is constrained only by
a condition that the triangle is closed. Such a triangle is a rigid
object and can be rotated into a predetermined position by an overall
conjugation. This is why the irreducible connections on a
3-fibered Seifert manifold have no moduli.

The rigidity of the triangle is considerably decreased if all three
of its vertices belong to the same big circle (see Fig. 4), i.e. if
the images of all $x_{i}$ belong to the same $U(1)$ subgroup of
$SU(2)$, as it happens for a stationary phase point on a face. In this
case, say, a middle vertex can be infinitesimally shifted in the plane
perpendicular to the line of triangle, with the sizes of the sides of
triangle changing only to the second order in the shift. Thus the
operator $L_{-}$ has two zero modes, $\dim H^{1}=2$, however there are
still no moduli, because an obstruction prevents an extension of those
modes to a 1-parameter family of flat connections.

The two zero modes form a 2-dimensional representation of $U(1)$ which
is a symmetry group of the reducible flat connections. This means that
the modes are gauge equivalent. However a procedure of dropping the
zero modes of $\Delta$ and $L_{-}$ from the determinants of
eq.~(\ref{2.01}) amounts to neglecting the global $U(1)$ gauge
transformations (see Appendix). The integration in path
integral~(\ref{2}) includes the directions along both zero modes. The
exponent corresponding to these directions has no quadratic terms, the
qubic terms are prohibited by the $U(1)$ symmetry. Therefore the
dominating term in the exponent is generally of the fourth order in
coordinates along the zero modes. Each direction contributes an
integral~(\ref{2.9}) which results in the formula~(\ref{2.10}) for
th factor $C_{1}$. Since in our case
$\dim H^{0}=1,\; \dim H^{1}=2,\;\dim X=0$ we see that eq.~(\ref{2.005})
predicts an overall power of $K$ to be equal to zero in agreement
with the surgery asymptotics~(\ref{4.3.1}).

\noindent
\underline{Point on an Edge}
\nopagebreak

Suppose that in addition to the previous conditions, one of the
phases $\phi_{i}$ is equal to zero or $\frac{1}{2}$.
This means that $\alpha^{({\rm st})}_{i}$ equals either 0 or $K$, so
that a stationary phase point belongs to an edge of the
tetrahedron~(\ref{3.3}).

Let, for example, $\phi_{1}=\alpha_{1}^{({\rm st})}=u_{1}=0$. Then the
image of $x_{1}$ in $SU(2)$ is the identity matrix. The triangle
$x_{1},x_{2}x_{1},x_{3}x_{2}x_{1}=1$ is even more degenerate, becasue
its first side has shrunk. The rigidity of the construction is
restored, the zero modes of $L_{-}$ disappeared, $\dim H^{1}=0$.
Since $\dim H^{0}=1$, then according to eq.~(\ref{2.005}) we expect a
contribution to be proportional to $k^{-1/2}$. Indeed, the
approximation~(\ref{4.3.1}) breaks down and the first subleading term
in eq.~(\ref{S.4}) contributes
\begin{equation}
i\frac{e^{2\pi i\lambda}}{\sqrt{2KH}}\frac{1}{p_{1}}
\left(\prod_{i=2}^{3}2\sin(2\pi\phi_{i})\right)
\exp2\pi i K\left(\sum_{i=1}^{3}\frac{r_{i}}{p_{i}}n_{i}^{2}+
\frac{H}{P}\lambda^{2}\right).
\label{4.3.2}
\end{equation}
This expression is very similar to eq.~(\ref{4.2.6}). We easily
recognize the same construction blocks assuming that now
\begin{equation}
\tau_{R}^{1/2}=\frac{1}{\sqrt{H}}\frac{1}{p_{1}}
\left|\prod_{i=2}^{3}2\sin(2\pi\phi_{i})\right|.
\label{4.3.3}
\end{equation}

\noindent
\underline{Point on a Vertex}
\nopagebreak

This is the case when the image of $\pi_{1}$ is a subgroup of the
center of $SU(2)$, that is,
$c_{0},\phi_{1},\phi_{2},\phi_{3}=0,\frac{1}{2}$. Let us take a
particular case of a trivial connection, for which
$c_{0}=\phi_{1}=\phi_{2}=\phi_{3}=0$. The Chern-Simons invariant is
zero, the square root of the Reidemeister torsion is known to be (see,
e.g.~\cite{FG})
\begin{equation}
\tau_{X}^{1/2}=H^{-\frac{\dim G}{2}}=H^{-3/2}
\label{4.3.4}
\end{equation}
The group H is the whole $SU(2)$, its volume in the proper
normalization is $1/(\sqrt{2}\pi)$ (see eq.~(\ref{2.3.14})). We also
know that $\dim H^{0}=3,\;\dim H^{1}=0$. As for the phase factor in
eq.~(\ref{2.005}), it is shown in~\cite{FG} that for the trivial
connection
\begin{equation}
\exp-\frac{i\pi}{4}\left[2I_{a}+3(1+b^{1})+\dim H^{0}+\dim
H^{1}\right]=1.
\label{4.3.5}
\end{equation}
Therefore, according to eq.~(\ref{2.005}), the 1-loop contribution of
the trivial connection should be
\begin{equation}
Z=\sqrt{2}\pi(KH)^{-3/2}.
\label{4.3.6}
\end{equation}
We get the same expression from the surgery calculus if we take the
term with $j=1$ in eq.~(\ref{S.5}).

\noindent
\underline{2-Loop Correction}
\nopagebreak

Let us use eq.~(\ref{S.5}) to calculate the next subleading correction
to the formula~(\ref{4.3.6})\footnote{I am indebted to D. Freed for
turning my attention to this calculation.}.
In other words, we are looking for a 2-loop term $S_{2}$ as defined by
eq.~(\ref{2.4}) (note, however, that we are using now $K=k+2$ instead
of $k$ as an expansion parameter). One obvious source of $S_{2}$ is
the 2-loop phase $-i\frac{\pi}{2}\phi$ (see eqs.~(\ref{S.2}) or
(\ref{4.1.13})). The other source is the $j=2$ term in the asymptotic
series of (\ref{S.5}). Actually, we have to take a logarithm of that
series to bring it to the form~(\ref{2.4}). At the 2-loop level of
approximation this amounts to dividing the $j=2$ term by the leading
$j=1$ term. Since
\begin{equation}
\left.\left[\partial_{\epsilon}^{(4)}
\frac{\prod_{i=1}^{3}2i\sin\left(\frac{\epsilon}{p_{i}}\right)}
{2i\sin\epsilon}\right]\right|_{\epsilon=0}=
\frac{16}{P}\left[\sum_{i=1}^{3}p_{i}^{-2}-1\right],
\label{C.1}
\end{equation}
then the whole 2-loop correction $S_{2}$ is
\begin{equation}
S_{2}=i\frac{\pi}{2}\left[\frac{P}{H}
\left(\sum_{i=1}^{3}p_{i}^{-2}-1\right)-
3{\rm sign}\left(\frac{H}{P}\right)-
\sum_{i=1}^{3}\left(12s(q_{i},p_{i})-\frac{q_{i}}{p_{i}}
\right)\right]=6\pi\lambda_{CW},
\label{C.2}
\end{equation}
here $\lambda_{CW}$ is a Casson-Walker invariant (see
eq.~(\ref{4.1.14})).

%*****************************************************
\section{A Large $k$ Limit of the Invariants of General Seifert
Manifolds}
%*****************************************************

%*****************************************************
\subsection{A Bernstein-Gelfand-Gelfand Resolution and Verlinde
Numbers}
\label{*5.1}
%*****************************************************

A calculation of Witten's invariant for a general Seifert manifold
can proceed along the same lines as that for the 3-fibered one,
described in section~\ref{*3}. We will try again to convert the sums
in eq.~(\ref{2.2.7}) into the integrals over the $n$-dimensional
half-spaces. We need a representation for Verlinde numbers
$N_{\alpha_{1},\ldots,\alpha_{n}}$ similar to that of
subsection~\ref{*3.1}. We will do this with the help of the
Bernstein-Gelfand-Gelfand resolution, which presents a representation
space of a Lie group $G$ as a cohomology over a complex of certain
vector spaces (see e.g. a review~\cite{BMP} and references therein).

We introduce the following notation. $\Delta$ with various subscripts
will denote the sets of weights of $G$ coming with multiplicities. In
other words, the elements of $\Delta$ are pairs $(v,m)$, where $v$ is
a weight and $m\in{\bf Z}$ is its multiplicity. The weights form an
abelian group. Consider its group algebra ${\cal A}$ with
the coefficients
in ${\bf Z}$. There is a one-to-one correspondence between the sets
$\Delta$ and the elements of ${\cal A}$:
\begin{equation}
\Delta\longleftrightarrow\sum_{(v,m)\in\Delta}mv.
\label{5.1.01}
\end{equation}
We define the sums and products of the sets $\Delta$ which parallel
the operations in ${\cal A}$. The sum $\Delta_{1}+\Delta_{2}$
consisits of weights belonging to either of the sets
$\Delta_{1},\Delta_{2}$ and coming with the multiplicities which are
sums of their multiplicities in $\Delta_{1}$ and $\Delta_{2}$. To
build a product $\Delta_{1}\circ \Delta_{2}$ we take all the pairs of
weights $v_{1}\in\Delta_{1},\;v_{2}\in\Delta_{2}$. Their sums
$v_{1}+v_{2}$ appear in the product $\Delta_{1}\circ \Delta_{2}$ with
multiplicities $m_{1}m_{2}$. If the same weight $v$ appears more than
once as a sum $v_{1}+v_{2}$, then we add all its multiplicities
in order to account
for the similar terms. A sum $\sum_{v\in\Delta}F(v)$ is a
shorthand for $\sum_{(v,m)\in\Delta}mF(v)$, in the same way a
product $\prod_{v\in\Delta}F(v)$ is equivalent to
$\prod_{(v,m)\in\Delta}F^{m}(v)$.

Consider a representation space $V_{\Lambda}$ of a Lie
group $G$ with the shifted highest weight $\Lambda$ (we remind that a
highest weight of $V_{\Lambda}$ is
$\Lambda-\rho,\;\;\rho=\frac{1}{2}\sum_{\lambda_{i}\in\Delta_{+}}
\lambda_{i},\;\;\Delta_{+}$ is a set of positive roots of $G$.
A Weyl formula for the character of $V_{\Lambda}$ as a function of the
element $x$ of a Cartan subalgebra, is a ratio
\begin{equation}
\chi_{\Lambda}(x)\equiv\sum_{v\in\Delta_{\Lambda}}e^{iv\cdot x}=
\frac{\sum_{w\in W}(-1)^{|w|}e^{i\left[w(\Lambda)-\rho\right]\cdot x}}
{\prod_{\lambda_{i}\in\Delta_{+}}\left(1-e^{-i\lambda_{i}\cdot
x}\right)}.
\label{5.1.1}
\end{equation}
Here $\Delta_{\Lambda}$ is a set of weights of $V_{\Lambda}$, $W$ is a
Weyl group and $|w|$ is a number of elementary Weyl reflections (mod
2) whose product is equal to $w$.

An individual term in the formula~(\ref{5.1.1}) can be presented as a
sum
\begin{equation}
(-1)^{|w|}\frac{e^{i\left[w(\Lambda)-\rho\right]\cdot x}}
{\prod_{\lambda_{i}\in\Delta_{+}}
\left(1-e^{-i\lambda_{i}\cdot x}\right)}=
(-1)^{|w|}\sum_{v\in\Delta^{(1)}_w(\Lambda)}
e^{iv\cdot x},
\label{5.1.2}
\end{equation}
here a set $\Delta_{\Lambda}^{(1)}$ contains all the weights of the
form
\begin{equation}
v=\Lambda-\rho-\sum_{i}n_{i}\lambda_{i},\;n_{i}\geq 0
\label{5.1.3}
\end{equation}
with multiplicity 1 (in fact, many weights may ultimately have a
bigger multiplicity, because generally the positive roots
$\lambda_{i}$ are not linearly independent).
It follows from eqs.~(\ref{5.1.1}) and~(\ref{5.1.2}) that
\begin{equation}
\Delta_{\Lambda}=\sum_{w\in W}(-1)^{|w|}\Delta_{w(\Lambda)}.
\label{5.1.4}
\end{equation}
The infinite dimensional modules of the Bernstein-Gelfand-Gelfand
resolution consist of vectors with the weights and multiplicities of
$\Delta^{(1)}_{w(\Lambda)}$.

The ``classical'' Verlinde numbers appear in the decomposition of the
tensor product
\begin{equation}
V_{\Lambda_{1}}\otimes V_{\Lambda_{2}}=
\sum_{\Lambda_{3}}N_{\Lambda_{1}\Lambda_{2}}^{\Lambda_{3}}
V_{\lambda_{3}}.
\label{5.1.04}
\end{equation}
A product of characters decomposes as
\begin{equation}
\chi_{\Lambda_{1}}(x)\chi_{\Lambda_{2}}(x)=
\sum_{\Lambda_{3}\in\Delta_{\Lambda_{1},\Lambda_{2}}}
\chi_{\Lambda_{3}}(x),
\label{5.1.5}
\end{equation}
here $\Delta_{\Lambda_{1},\Lambda_{2}}$ is a set of shifted highest
weights of all representations appearing in the
decomposition~(\ref{5.1.04}) and coming with the multiplicities
$N_{\Lambda_{1}\Lambda_{2}}^{\Lambda_{3}}$. Note that we raised a
third index of Verlinde numbers. The indices are raised and lowered by
the metric $N_{\Lambda_{1}\Lambda_{2}}$ which is equal to 1 if
$V_{\Lambda_{1}}$ and $V_{\Lambda_{2}}$ are conjugate representations,
and is zero otherwise. For $G=SU(2)$ there is no distinction between
the upper and lower indices since
$N_{\alpha_{1}\alpha_{2}}=\delta_{\alpha_{1}\alpha_{2}}$.

If we use the r.h.s of eq.~(\ref{5.1.1}) for $\chi_{\Lambda_{1}}$ and
$\chi_{\Lambda_{3}}$, and the middle expression of eq.~(\ref{5.1.1})
for $\chi_{\Lambda_{2}}$, then we see that
\begin{equation}
\sum_{w\in W}\sum_{v\in\Delta_{\Lambda_{2}}}(-1)^{|w|}
e^{i\left[w(\Lambda_{1})+v\right]\cdot x}=
\sum_{w\in W}\sum_{\Lambda_{3}\in\Delta_{\Lambda_{1},\Lambda_{2}}}
(-1)^{|w|}e^{iw(\Lambda_{3})\cdot x}.
\label{5.1.6}
\end{equation}
Let us denote by $\Delta_{W(\Lambda)}$ a set containing all the
weights $w(\Lambda),\;w\in W$ with multiplicities $(-1)^{|w|}$. Then
it is easy to translate eq.~(\ref{5.1.6})
into a statement about sets:
\begin{equation}
\Delta_{W(\Lambda_{1})}\circ
\Delta_{\Lambda_{2}}=
\sum_{\Lambda_{3}\in\Delta_{\Lambda_{1},\Lambda_{2}}}
\Delta_{W(\Lambda_{3})}.
\label{5.1.7}
\end{equation}
Finally applying eq.~(\ref{5.1.4}) to $\Delta_{\Lambda_{2}}$ we get
\begin{equation}
\sum_{w\in W}(-1)^{|w|}
\Delta_{W(\Lambda_{1})}\circ
\Delta_{w(\Lambda_{2})}^{(1)}=
\sum_{\Lambda_{3}\in\Delta_{\Lambda_{1},\Lambda_{2}}}
\Delta_{W(\Lambda_{3})},
\label{5.1.8}
\end{equation}
or equivalently,
\begin{equation}
\sum_{w_{1,2}\in W}(-1)^{|w_{1}|+|w_{2}|}
\Delta_{w_{1}(\Lambda_{1})+w_{2}(\Lambda_{2})}^{(1)}=
\sum_{\Lambda_{3}\in\Delta_{\Lambda_{1},\Lambda_{2}}}
\Delta_{W(\Lambda_{3})}.
\label{5.1.9}
\end{equation}

It is easy to generalize this relation to a tensor product of $n-1$
vector spaces:
\begin{equation}
\sum_{w_{1},\ldots,w_{n-1}\in W}
(-1)^{|w_{1}|+\cdots+|w_{n-1}|}
\Delta_{w_{1}(\Lambda_{1})+\cdots+w_{n-1}(\Lambda_{n-1})}^{(n-2)}
=\sum_{\Lambda_{n}\in\Delta_{\Lambda_{1},\ldots,\Lambda_{n-1}}}
\Delta_{W(\Lambda_{n})},
\label{5.1.10}
\end{equation}
here $\Delta_{\Lambda_{1},\ldots,\Lambda_{n-1}}$ is a set of weights
$\Lambda_{n}$ taken with multiplicities
$N_{\Lambda_{1}\ldots\Lambda_{n-1}}^{\Lambda_{n}}$, while
$\Delta_{\Lambda}^{(n)}$ contains all the weights
\begin{equation}
v=\Lambda-n\rho-\sum_{i}\sum_{j=1}^{n}n_{i,j}\lambda_{i},\;\;
n_{i,j}\geq 0
\label{5.1.11}
\end{equation}
coming with multiplicities 1 (before the counting of similar terms).
There are $\left(\begin{array}{c}n-1\\m+n-1\end{array}\right)$ ways in
which a number $m$ can be represented as a sum of $n$ nonnegative
numbers. Therefore we can say that $\Delta_{\Lambda}^{(n)}$ consists
of the weights
\begin{equation}
v=\Lambda-n\rho-\sum_{i}n_{i}\lambda_{i},
\label{5.1.12}
\end{equation}
coming with multiplicities
$\prod_{i}\left(
\begin{array}{c}n-1\\n_{i}+n-1\end{array}\right)$.

We need an analog of eq.~(\ref{5.1.10}) for the case of affine Lie
algebra (or a quantum group $G_{q}$). The affine Weyl group
$\tilde{W}$ is a semidirect product of $W$ and an abelian group  $T$
of translations by the elements of the root lattice multiplied by $K$.
We can not simply substitute $\tilde{W}$ for $W$ in
eq.~(\ref{5.1.10}), because the previous reasoning does not quite
apply to the case of affine algebras (e.g. eq.~(\ref{5.1.5}) is no
longer valid). Still it turns out that a simple modification of
eq.~(\ref{5.1.10}) makes it work for affine algebras or quantum
groups:
\begin{equation}
\sum_{t\in T}
\sum_{w_{1},\ldots,w_{n-1}\in W}
(-1)^{|w_{1}|+\cdots+|w_{n-1}|}
\Delta_{t(w_{1}(\Lambda_{1})+\cdots+w_{n-1}(\Lambda_{n-1}))}^{(n-2)}=
\sum_{\Lambda_{n}\in\Delta_{\Lambda_{1},\ldots,\Lambda_{n-1}}}
\Delta_{\tilde{W}(\Lambda_{n})},
\label{5.1.13}
\end{equation}
here $\Delta_{\tilde{W}(\Lambda)}$ is a set of weights
$w(\Lambda),\;w\in\tilde{W}$ coming with multiplicities $(-1)^{|w|}$,
$|w|$ counts only the number of reflections\footnote{
The same equation can be derived directly from Verlinde's formula
\begin{displaymath}
N_{\Lambda_{1},\ldots,\Lambda_{n}}=
\sum_{\Lambda\in\Delta}\left(\prod_{i=1}^{n}
S_{\Lambda\Lambda_{i}}\right)/S_{\rho\Lambda}^{n-2}
\end{displaymath}
($\Delta$ being a set of integrable highest weights), by expanding its
denominator in a geometric series similar to that of eq.~(\ref{5.1.2})
and performing a Poisson resummation on $\Lambda$. In fact, $\Lambda$
plays a role very similar to $c$ in eq.~(\ref{5.2.2}).

A generalized Verlinde's formula
\begin{equation}
N_{\Lambda_{1},\ldots,\Lambda_{n}}^{(g)}=
\sum_{\Lambda\in\Delta}\left(\prod_{i=1}^{n}
S_{\Lambda\Lambda_{i}}\right)/S_{\rho\Lambda}^{n-2+2g}
\label{F1}
\end{equation}
for the number of conformal blocks on a $g$-handled Riemann surface
$\Sigma_{g}$ with $n$ primary fields ${\cal V}_{\Lambda_{i}}$, allows
us to generalize the results of our calculations to the case of a
Seifert manifold constructed by a surgery on circles in
$S^{1}\times\Sigma_{g}$. Eq.~(\ref{F1}) suggests that the presence
of handles can be accounted for by a simple substitution
$n\rightarrow n+2g$ in eqs.~(\ref{5.1.14})-(\ref{5.1.161})
and~(\ref{5.1.20}). Thus only a multiplicity factor $N_{n}(x)$ is
affected (it is substituted by $N_{n+2g}$), while other quantities,
such as Chern-Simons action of flat connections, remain unchanged, so
that eq.~(\ref{4.1.5}) is still valid in agreement with \cite{A}. We
hope to discuss this subject further in a forthcoming paper.}.

The l.h.s. of eq.~(\ref{5.1.14}) consists of all the weights
\begin{equation}
v=\sum_{i=1}^{n-1}w_{i}(\Lambda_{i})-(n-2)\rho-
\sum_{i}n_{i}\lambda_{i}+K\sum_{i}m_{i}\lambda_{i},\;\;
n_{i},m_{i}\in{\bf Z},\;n_{i}\geq0
\label{5.1.14}
\end{equation}
coming with multiplicities
\begin{equation}
(-1)^{\sum_{i=1}^{n-1}|w_{i}|}\prod_{i}\left(
\begin{array}{c}n-3\\n_{i}+n-3\end{array}\right).
\label{5.1.15}
\end{equation}
For a given set of numbers $m_{i}$ and Weyl reflections $w_{i}$, these
vectors form a half-space similar to that of eq.~(\ref{3.15}). The
r.h.s. of eq.~(\ref{5.1.13}) consists of the highest weights
$\Lambda_{n}$ of integrable representations of affine Lie algebra
coming with multiplicities
$N_{\Lambda_{1}\ldots\Lambda_{n-1}}^{\Lambda_{n}}$ together with
all their images $w(\Lambda_{n}),\;w\in\tilde{W}$, whose
multiplicities have an extra factor $(-1)^{|w|}$. In other words,
the r.h.s. of eq.~(\ref{5.1.13}) consists of all the weights
$\Lambda_{n}$ coming with the multiplicities
$\tilde{N}_{\Lambda_{1}\ldots\Lambda_{n-1}}^{\Lambda_{n}}$ which are
Verlinde numbers
$N_{\Lambda_{1}\ldots\Lambda_{n-1}}^{\Lambda_{n}}$ extended to all the
weights of $G$ by the affine Weyl group: the extended numbers
$\tilde{N}_{\Lambda_{1}\ldots\Lambda_{n-1}}^{\Lambda_{n}}$ are
invariant under the shifts of $T$ and they are antisymmetric under the
Weyl reflections. Since the matrices $\tilde{M}$ of eq.~(\ref{2.2.7})
exhibit the same properties under the action of $\tilde{W}$, we can
use the extended Verlinde numbers as defined by
eqs.~(\ref{5.1.13})-(\ref{5.1.15}) in order to extend the sums in
eq.~(\ref{2.2.7}) from the integrable highest weights to the whole
weight lattice of $G$ and to transform it into a sum over the
``half-spaces''~(\ref{5.1.14}),(\ref{5.1.15}).

\noindent
\underline{The Case of $SU(2)$}
\nopagebreak

Let us study specifically the case of $G=SU(2)$. We remind that the
variables $\alpha$ play the role of shifted highest weights, $\rho=1$
and the only positive root is equal to 2. Thus
according to eq.~(\ref{5.1.13}), we can drop Verlinde numbers
$N_{\alpha_{1}\cdots\alpha_{n}}$ from eq.~(\ref{2.2.7}) if we take
the sum there over all the n-dimensional vectors
$(\alpha_{1},\ldots,\alpha_{n})$ satisfying an equation
\begin{equation}
\sum_{i=1}^{n}\nu_{i}\alpha_{i}=(n-2)+2m+2Kl
\label{5.1.16}
\end{equation}
and coming with multiplicities
\begin{equation}
\left(\begin{array}{c}n-3\\m+n-3\end{array}\right)
\prod_{i=1}^{n}\nu_{i}.
\label{5.1.161}
\end{equation}
Here
\begin{equation}
\nu_{1},\ldots,\nu_{n-1}=\pm 1,\;\;\nu_{n}=-1,\;\;
m,l\in{\bf Z},\;\;m\geq 0.
\label{5.1.17}
\end{equation}
The multiplicity appears as a new factor in eq.~(\ref{2.2.7}) taking
the place of Verlinde numbers.

We can make a substitution
\begin{equation}
m=\frac{1}{2}(x-n+2),
\label{5.1.18}
\end{equation}
so that eq.~(\ref{5.1.16}) transforms into
\begin{equation}
\sum_{i=1}^{n}\nu_{i}\alpha_{i}=x+2Kl
\label{5.1.19}
\end{equation}
and the multiplicity factor is
\begin{equation}
N_{n}(x)=-\frac{\prod_{i=1}^{n}\nu_{i}}{2^{n-3}(n-3)!}
\prod_{\stackrel{-n+4\leq i\leq n-4}{i+n\;{\rm even}}}(x-i).
\label{5.1.20}
\end{equation}
The substitution~(\ref{5.1.18}) requires that $x-n$ is even and $x\geq
n-2$. In fact we
may demand only that $x\geq 0$, because the fixed parity
together with the last factor of eq.~(\ref{5.1.20}) eliminates all
possible extra values of $x$.

%*****************************************************
\subsection{A Contribution of Irreducible Flat Connections}
\label{*5.2}
%*****************************************************

We have to calculate a sum
\begin{equation}
\sum_{\stackrel{x>0}{x-n\;{\rm even}}}
\sum_{\alpha_{i}:\;\sum_{i=1}^{n}\nu_{i}\alpha_{i}=x+2Kl}
N_{n}(x)\prod_{i=1}^{n}\tilde{M}^{(p_{i},q_{i})}_{\alpha_{i}1}.
\label{5.2.1}
\end{equation}
Similarly to subsection~\ref{*3.1} we turn a sum over $\alpha_{i}$
into an integral by extending the sum over $n$ in eq.~(\ref{2.3.2}) to
all integer numbers. We take care of a condition~(\ref{5.1.19}) by
adding a factor
\begin{equation}
\int_{-\infty}^{+\infty}dc\,\exp2\pi
ic\,\left(x+2Kl-\sum_{i=1}^{3}\nu_{i}\alpha_{i}\right)
\label{5.2.2}
\end{equation}
 to eq.~(\ref{5.2.1}). Another familiar factor
\begin{equation}
\frac{1}{2}\sum_{\lambda=0,\frac{1}{2}}\exp2\pi i\lambda
\left(\sum_{i=1}^{n}\alpha_{i}+n\right)
\label{5.2.3}
\end{equation}
guarantees together with the factor~(\ref{5.2.2}) that $x-n$ is even.
Finally since the factor~(\ref{5.2.2}) makes $x$ an integer, we
can take an integral over $x$ rather than a sum:
\begin{eqnarray}
Z_{\nu_{1},\ldots,\nu_{n};l}&=&
Z_{f}\frac{1}{2}\sum_{\lambda=0,\frac{1}{2}}e^{2\pi i\lambda n}
\int_{0}^{\infty}dx\,N_{n}(x)\int_{-\infty}^{+\infty}dc\,
\exp 2\pi ic(x+2Kl)
\nonumber\\
&&\times
\prod_{i=1}^{n}\sum_{n_{i}\in{\bf Z}}
\sum_{\mu_{i}=\pm 1}\mu_{i}
\int_{-\infty}^{+\infty}d\alpha_{i}\,i
%\nonumber\\
%&&\times
\frac{{\rm sign}(q_{i})}{\sqrt{2K|q_{i}|}}
e^{-i\frac{\pi}{4}\Phi\left(M^{(p_{i},q_{i})}\right)}
\nonumber\\
&&\times
\exp\frac{i\pi}{2Kq_{i}}
\left[p_{i}\alpha_{i}^{2}
+2\alpha_{i}(2K(\tilde{n}_{i}+\nu_{i}q_{i}c)+\mu_{i})+
s_{i}(2Kn_{i}+\mu_{i})^{2}\right],
\label{5.2.4}
\end{eqnarray}
here $Z_{f}$ is a framing correction given by eq.~(\ref{3.3.4}) with a
substitution $\sum_{i=1}^{3}\longrightarrow\sum_{i=1}^{n}$.

We fix the numbers $n_{i}$ in order to study a contribution of a
particular point~(\ref{3.17}). After integrating over $\alpha_{i}$
and $c$, a partition function becomes a product of two factors
$Z_{1}$ and $Z_{2}$:
\begin{eqnarray}
&&Z_{\nu_{1},\ldots,\nu_{n};l}=Z_{1}Z_{2}
\label{5.2.5}
%\nonumber
\\
Z_{1}&=&e^{i\frac{3\pi}{4}{\rm sign}\left(\frac{H}{P}\right)}
\left(\prod_{i=1}^{n}{\rm sign}\,p_{i}\right)
\exp-\frac{i\pi}{2K}\left[3{\rm sign}\left(\frac{H}{P}\right)+
\sum_{i=1}^{n}\left(12s(q_{i},p_{i})-\frac{q_{i}}{p_{i}}\right)\right]
\label{5.2.6}
%\nonumber
\\
Z_{2}&=&\frac{e^{-i\frac{\pi}{4}{\rm sign}\left(\frac{H}{P}\right)}}
{\sqrt{2K|H|}}\frac{1}{2}\sum_{\lambda=0,\frac{1}{2}}
e^{2\pi i\lambda n}
\sum_{\mu_{1},\ldots,\mu_{n}=\pm 1}
\left[\prod_{i=1}^{n}\mu_{i}\exp2\pi iK
\left(\frac{r_{i}}{p_{i}}\tilde{n}_{i}^{2}-s_{i}q_{i}\lambda^{2}\right)
\right.
\nonumber\\
&&
\left.\times
\exp2\pi i\mu_{i}\left(\frac{r_{i}}{p_{i}}\tilde{n}_{i}
+s_{i}\lambda\right)\right]
%\nonumber\\
%&&\times
I(\nu_{1},\ldots,\nu_{n};\mu_{1},\ldots,\mu_{n};l),
\label{5.2.8}
\end{eqnarray}
here
\begin{eqnarray}
I(\nu_{1},\ldots,\nu_{n};\mu_{1},\ldots,\mu_{n};l)&=&
\int_{x_{0}}^{\infty}dx\,N_{n}(x-x_{0})
\exp\left[\frac{i\pi}{2K}\frac{P}{H}
\left(x-\sum_{i=1}^{n}\nu_{i}\frac{\mu_{i}}{p_{i}}\right)^{2}\right]
\label{5.2.9}\\
x_{0}&=&2Kl-\sum_{i=1}^{n}\nu_{i}\alpha_{i}^{({\rm st})},
\label{5.2.10}
\end{eqnarray}
and we remind that $\alpha_{i}^{({\rm st})}$ are defined by
eq.~(\ref{3.17}). We made a change of variables
$x\longrightarrow x-x_{0}$ in deriving eq.~(\ref{5.2.9}).

The integral~(\ref{5.2.9}) is gaussian apart from a polynomial
factor $N_{n}(x-x_{0})$. This integral is similar to the
integral~(\ref{3.2.7}) and should be treated in a similar way. Here we
are concerned with a contribution of a stationary phase point $x=0$,
which contributes to the integral~(\ref{5.2.9}) if $x_{0}<0$.

Consider a specific point $\alpha_{i}^{({\rm st})}$ of
eq.~(\ref{3.17}). As we know from subsection~\ref{*3.1}, we should
limit our attention to the points belonging to the fundamental cube
\begin{equation}
0\leq \alpha_{i}^{({\rm st})}<K.
\label{5.2.12}
\end{equation}
We have to determine a set ${\cal S}$ of arrays
$(\nu_{1},\ldots\nu_{n-1};l)$ for which a stationary phase point
$\alpha_{i}^{({\rm st})} $ contributes to the integral~(\ref{5.2.9}).
Then we should substitute a sum
\begin{equation}
N_{n}^{({\rm tot})}(x)=\sum_{\cal S}N_{n}(x+\sum_{i=1}^{n}
\nu_{i}\alpha_{i}^{({\rm st})}-2Kl)
\label{5.2.13}
\end{equation}
instead of $N_{n}(x-x_{0})$ in eq.~(\ref{5.2.9}) and extend the
integral over $x$ to $\int_{-\infty}^{+\infty}$ in order to get a full
contribution of the point $\alpha_{i}^{({\rm st})}$ to the partition
function~(\ref{5.2.5}).

If we compose the set ${\cal S}$ of all arrays
$(\nu_{1},\ldots\nu_{n-1};l)$
for which
\begin{equation}
2Kl-\sum_{i=1}^{n}\nu_{i}\alpha_{i}^{({\rm st})}\leq0,
\label{5.2.14}
\end{equation}
then the sum~(\ref{5.2.13}) will contain infinitely many terms.
However most of these terms will cancel each other. Therefore we
propose another procedure that will express $N_{n}^{({\rm tot})}$ as a
finite sum. Suppose for simplicity that $\alpha_{n}^{({\rm st})}\neq0$.
Consider a line in the $\alpha$-space
\begin{equation}
\alpha_{i}(t)=t\alpha_{i}^{({\rm st})},\;i=1,\ldots,n-1;\;\;
\alpha_{n}(t)=\alpha_{n}^{({\rm st})}.
\label{5.2.15}
\end{equation}
Obviously, $\alpha_{i}(1)=\alpha_{i}^{({\rm st})}$. Remember now that
$N_{n}^{({\rm tot})}$ is a Verlinde number. Therefore $N_{n}^{({\rm
tot})}=0$ for $\alpha_{i}(0)$. This means that all the terms in
eq.~(\ref{5.2.13}) cancel each other and we may drop them altogether.
As $t$ starts to grow, suppose that for some value $t_{*}$
\begin{equation}
2Kl-\sum_{i=1}^{n}\nu_{i}\alpha_{i}(t_{*})=0.
\label{5.2.16}
\end{equation}
If for $t>t_{*}$ the l.h.s. of eq.~(\ref{5.2.16}) is negative, then by
passing $t=t_{*}$ we gained a contribution of the array
$(\nu_{1},\ldots,\nu_{n-1};l)$ to eq.~(\ref{5.2.13}) and the
corresponding term should be added there. If, however, for $t>t_{*}$
the l.h.s. of eq.~(\ref{5.2.16}) is positive, then we lost the
contribution of the array $(\nu_{1},\ldots,\nu_{n-1};l)$, and the
corresponding term should be subtracted. As a result,
\begin{equation}
N_{n}^{({\rm tot})}(x)=-\sum_{{\cal S}^{\prime}}
{\rm sign}\left(\sum_{i=1}^{n-1}\nu_{i}\alpha_{i}\right)
N_{n}\left(x+\sum_{i=1}^{n}\nu_{i}\alpha_{i}^{({\rm st})}-2Kl\right),
\label{5.2.17}
\end{equation}
here a set ${\cal S}^{\prime}$ consists of all arrays
$(\nu_{1},\ldots,\nu_{n-1};l)$ such that for some $t_{*}\in [0,1]$
eq.~(\ref{5.2.16}) is satisfied

Since $\alpha_{i}^{({\rm st})}$ are proportional to $K$, the function
$N_{n}^{({\rm tot})}(x)$ is a polynomial in $K$ and $x$:
\begin{equation}
N_{n}^{({\rm tot})}(x)=\sum_{j=0}^{n-3}C_{j}K^{n-3-j}x^{j}.
\label{5.2.18}
\end{equation}
A Verlinde number
$N_{\alpha_{1}^{({\rm st})}\ldots\alpha_{n}^{({\rm st})}}=
N_{n}^{({\rm tot})}(0)=C_{0}$ is a number of the WZW conformal
blocks of $n$ primary fields ${\cal V}_{\alpha_{i}^{({\rm st})}}$ on a
sphere\footnote{The numbers $\alpha_{i}^{({\rm st})}$ are not
necessarily integer, so in fact, we should take the closest integer
numbers. This does not change a conclusion that $C_{0}$ is the volume
of the moduli space}. This number to the leading power in $K$ is
proportional to the volume of the moduli space of flat connections on
a sphere with $n$ punctures, the holonomies around which are fixed by
eq.~(\ref{4.1.3}). This moduli space coincides with a
connected component of the moduli space of flat connections on the
Seifert manifold, for which the map~(\ref{2.2}) is determined by
eq.~(\ref{4.1.3}). Therfore the coefficient $C_{0}$ is equal to the
volume of that component of the moduli space (calculated with the
proper measure).

Let us substitute eq.~(\ref{5.2.18}) into the integral~(\ref{5.2.9})
modified in order to get the contribution of the stationary phase
point $\alpha_{i}^{({\rm st})}$:
\begin{equation}
I=\sum_{j=0}^{n-3}C_{j}K^{n-3-j}
\int_{-\infty}^{+\infty}dx\,
\left(x+\sum_{i=1}^{n}\nu_{i}\frac{\mu_{i}}{p_{i}}\right)^{j}
\,\exp\left(\frac{i\pi}{2K}\frac{P}{H}x^{2}\right).
\label{5.2.19}
\end{equation}
The dominant contribution comes from the term with $j=0$. It is
proportional to $K^{n-3}$:
\begin{equation}
I\approx e^{i\frac{\pi}{4}{\rm sign}\left(\frac{H}{P}\right)}
\sqrt{\frac{2K|H|}{|P|}}C_{0}K^{n-3},
\label{5.2.20}
\end{equation}
so that the whole 1-loop partition function contribution coming form
the point~(\ref{3.17}) is (for $p_{1},p_{2},p_{3},H>0$)
\begin{eqnarray}
Z&\approx&\frac{1}{2}e^{i\frac{3\pi}{4}}\exp-\frac{i\pi}{2K}
\left[3+\sum_{i=1}^{n}\left(12s(q_{i},p_{i})-\frac{q_{i}}{p_{i}}
\right)\right]K^{n-3}C_{0}
\nonumber\\
&&\times\sum_{\lambda=0,\frac{1}{2}}e^{2\pi i\lambda n}
\prod_{i=1}^{n}
\exp2\pi iK\left(\frac{r_{i}}{p_{i}}\tilde{n}_{i}^{2}
-s_{i}q_{i}\lambda^{2}\right)\frac{2i}{\sqrt{p_{i}}}
\sin2\pi\left(\frac{r_{i}}{p_{i}}\tilde{n}_{i}+s_{i}\lambda\right).
\label{5.2.21}
\end{eqnarray}
Comparing this expression with eq.~(\ref{2.005}) we note that $\dim
H^{1}=2(n-3)$, hence the factor $K^{n-3}$. Also an integral over the
moduli space (which is included in the sum over flat connections in
eq.~(\ref{2.005})) produces its volume $C_{0}$.

In contrast to the results of subsection~\ref{*4.1}, the contribution
of the irreducible flat connection on an $n$-fibered Seifert manifold
($n\geq4$) contains higher loop corrections coming from the sum
in eq.~(\ref{5.2.19}). However the number of these corrections is
finite. The highest order correction is of the order of $K^{0}$, so
the number of loop corrections is equal to half the dimension of the
moduli space.

%*************************************************
\subsection{A Contribution of Reducible Flat Connections}
\label{*5.3}
%**************************************************

A contribution of a conditional stationary phase point~(\ref{3.18})
is easier to calculate, than that of an unconditional one
$\alpha^{(st)}$, because the former involves an integral only over one
half-space~(\ref{5.1.19}), to which boundary it belongs. We will use
an expression for $Z_{2}$ which is slightly different from that of
eq.~(\ref{5.2.4}) and is a generalization of eq.~(\ref{3.2.6}).
We express the multiplicity factor $N_{n}(x)$ of eq.~(\ref{5.1.20}) as
a derivative:
\begin{equation}
N_{n}(x)=-\frac{\prod_{i=1}^{n}\nu_{i}}{(n-3)!}
\left.\partial_{a}^{(n-3)}a^{\frac{1}{2}(x+n-4)}
\right|_{a=1}.
\label{5.3.01}
\end{equation}
We also shift the integration range of $x$ from $x\geq 0$ to $x\geq
4-n$. The contribution of the extra values of $x$ is killed by the
zeros of $N_{n}(x)$ (recall that $x$ is actually an even integer).
After a shift in the integration variable $x\rightarrow x+n-4$ we get
the following expression for $Z_{2}$ (the other factor $Z_{1}$ is
defined by eq.~(\ref{5.2.6})):
\begin{eqnarray}
Z_{2}&=&-\left[\prod_{i=1}^{n}e^{-i
\frac{\pi}{4}{\rm sign}(p_{i}q_{i})}
\exp\left(-\frac{i\pi}{2K}\frac{r_{i}}{p_{i}}\right)\right]
\sum_{\lambda=0,\frac{1}{2}}e^{2\pi i\lambda n}
\nonumber\\
&&\times
\sum_{m\in{\bf Z}}\int_{0}^{\infty}dx
\left[\left.\frac{1}{(n-3)!}\partial_{a}^{(n-3)}
a^{\frac{x}{2}}\right|_{a=1}\right]
\int_{-\infty}^{+\infty}dc\,e^{-2\pi ic(x+4-n+2Kl)}
\prod_{i=1}^{n}\sum_{\mu_{i}=\pm 1}\mu_{i}
\int_{-\infty}^{+\infty}\frac{d\alpha_{i}}{\sqrt{2K|q_{i}|}}
\nonumber\\
&&\times
\exp\frac{i\pi}{2Kq_{i}}\left[p_{i}\alpha_{i}^{2}-
2\nu_{i}\alpha_{i}(2Kn_{i}+2Kq_{i}(\lambda+m-c)+\mu_{i})+
s_{i}(2K(n_{i}+q_{i}m)+\mu_{i})^{2}\right]
\nonumber\\
&=&-\frac{e^{-i\frac{\pi}{4}{\rm sign}\left(\frac{H}{P}\right)}}
{\sqrt{2K|H|}}\exp 2\pi iK\left[\sum_{i=1}^{n}\frac{r_{i}}{p_{i}}
n_{i}^{2}+\frac{H}{P}c_{0}^{2}\right]
\nonumber\\
&&\times\sum_{\mu_{1},\ldots,\mu_{n}=\pm 1}
\left[\prod_{i=1}^{n}\mu_{i}\exp\left(2\pi i\mu_{i}
\frac{r_{i}n_{i}+c_{0}}{p_{i}}\right)\right]
\frac{1}{2}\sum_{\lambda=0,\frac{1}{2}}
\sum_{m\in{\bf Z}}I(m,\lambda),
\label{5.3.1}
\end{eqnarray}
here $c_{0}$ is defined by eq.~(\ref{3.19}), while
\begin{eqnarray}
I(m,\lambda)&=&\frac{e^{2\pi ic_{0}(n-4)}}{(n-3)!}
\int_{0}^{\infty}dx
\left[\left.\partial_{a}^{(n-3)}a^{\frac{x}{2}}\right|_{a=1}\right]
\nonumber\\
&&\times\exp\left[-2\pi ix(c_{0}+m+\lambda)+
\frac{i\pi}{2K}\frac{P}{H}\left(x-n+4-\sum_{i=1}^{n}
\frac{\mu_{i}}{p_{i}}\right)^{2}\right]
\nonumber\\
&=&\frac{e^{2\pi ic_{0}(n-4)}}{(n-3)!}
\sum_{j=0}^{\infty}\frac{1}{j!}(8\pi iK)^{-j}
\left(\frac{P}{H}\right)^{j}\partial_{\epsilon}^{(2j)}
\partial_{a}^{(n-3)}\left\{e^{2\pi i\epsilon\left(
\sum_{i=1}^{n}\frac{\mu_{i}}{p_{i}}+n-4\right)}
\right.\nonumber\\
&&\left.\left.\times\int_{0}^{\infty}dx\,a^{\frac{x}{2}}
\exp[-2\pi ix(c_{0}+m+\lambda+\epsilon)]\right\}\right|_{
\stackrel{a=1}{\epsilon=0}
}.
\label{5.3.2}
\end{eqnarray}

\noindent
\underline{General Reducible Connection}
\nopagebreak

A sum over $m$ and $\lambda$ converts an integral over $x$ in
eq.~(\ref{5.3.2}) into a sum over even $x$:
\begin{eqnarray}
\frac{1}{2}\sum_{\lambda=0,\frac{1}{2}}\sum_{m\in{\bf Z}}I(m)&=&
\frac{e^{2\pi ic_{0}(n-4)}}{(n-3)!}
\sum_{j=0}^{\infty}\frac{1}{j!}(8\pi iK)^{-j}
\left(\frac{P}{H}\right)^{j}
\nonumber\\
&&\times\partial_{\epsilon}^{(2j)}
\partial_{a}^{(n-3)}
\left.\left[e^{2\pi i\epsilon\left(\sum_{i=1}^{n}
\frac{\mu_{i}}{p_{i}}+n-4\right)}
\sum_{\stackrel{x\geq 0}{x-{\rm even}}}
a^{\frac{x}{2}}e^{-2\pi ix(c_{0}+\epsilon)}\right]
\right|_{\stackrel{a=1}{\epsilon=0}}
\nonumber\\
&=&\sum_{j=0}^{\infty}\frac{1}{j!}(8\pi iK)^{-j}
\left(\frac{P}{H}\right)^{j}\partial_{\epsilon}^{(2j)}
\left.\frac{e^{2\pi i\epsilon\sum_{i=1}^{n}\frac{\mu_{i}}{p_{i}}}}
{\left[2i\sin 2\pi(c_{0}+\epsilon)\right]^{n-2}}
\right|_{\epsilon=0}.
\label{5.3.3}
\end{eqnarray}
Therefore the total contribution of a general reducible connection is
\begin{eqnarray}
Z&=&-\frac{e^{i\frac{\pi}{2}{\rm sign}\left(\frac{H}{P}\right)}}
{\sqrt{2K|H|}}{\rm sign}(P)
\exp-\frac{i\pi}{2K}\left[3{\rm sign}\left(\frac{H}{P}\right)+
\sum_{i=1}^{n}\left(12s(q_{i},p_{i})-\frac{q_{i}}{p_{i}}\right)\right]
\nonumber\\
&&\times\exp 2\pi iK\left(\sum_{i=1}^{n}\frac{r_{i}}{p_{i}}n_{i}^{2}
+\frac{H}{P}c_{0}^{2}\right)
\sum_{j=0}^{\infty}\frac{1}{j!}(8\pi iK)^{-j}
\left(\frac{P}{H}\right)^{j}
\nonumber\\
&&\times\partial_{c}^{(2j)}\left.\left[
\frac{\prod_{i=1}^{n}2i\sin\left(2\pi
\frac{r_{i}n_{i}+c}{p_{i}}\right)}
{[2i\sin(2\pi c)]^{n-2}}\right]\right|_{c=c_{0}}.
\label{5.3.4}
\end{eqnarray}

\noindent
\underline{Special Reducible Connection}
\nopagebreak

If $c_{0}+m_{0}+\lambda_{0}=0$ for some values
$\lambda_{0}=0,\frac{1}{2},\;\;m_{0}\in{\bf Z}$, then a calculation of
$I(m_{0},\lambda_{0})$ has to be performed separately:
\begin{eqnarray}
I(m_{0},\lambda_{0})&=&\frac{e^{2\pi ic_{0}n}}{(n-3)!}
\int_{0}^{\infty}dx\left[\left.\partial_{a}^{(n-3)}a^{\frac{x}{2}}
\right|_{a=1}\right]\exp\left[\frac{i\pi}{2K}\frac{P}{H}
\left(x-n+4-\sum_{i=1}^{n}\frac{\mu_{i}}{p_{i}}\right)^{2}\right]
\nonumber\\
&=&\frac{e^{2\pi ic_{0}n}}{(n-3)!}
\left(\int_{0}^{\infty}-\int_{0}^{n-4+\sum_{i=1}^{n}
\frac{\mu_{i}}{p_{i}}}\right)\,dx\,
\left[\left.\partial_{a}^{(n-3)}
a^{\frac{1}{2}\left(x+n-4+\sum_{i=1}^{n}\frac{\mu_{i}}{p_{i}}
\right)}\right|_{a=1}\right]\exp\left(\frac{i\pi}{2K}
\frac{P}{H}x^{2}\right)
\nonumber\\
&=&\frac{e^{2\pi ic_{0}n}}{(n-3)!}\left\{
\int_{0}^{\infty}dx
\left[\left.\partial_{a}^{(n-3)}
a^{\frac{1}{2}\left(x+n-4+\sum_{i=1}^{n}\frac{\mu_{i}}{p_{i}}
\right)}\right|_{a=1}\right]\exp\left(\frac{i\pi}{2K}
\frac{P}{H}x^{2}\right)
\right.
\\
&&\left.-\sum_{j=0}^{\infty}\frac{1}{j!}(8\pi iK)^{-j}
\left(\frac{P}{H}\right)^{j}\partial_{\epsilon}^{(2j)}
\partial_{a}^{(n-3)}\left.\frac{
e^{2\pi i\epsilon\left(n-4+\sum_{i=1}^{n}\frac{\mu_{i}}{p_{i}}\right)}
-a^{\frac{1}{2}\left(n-4+\sum_{i=1}^{n}\frac{\mu_{i}}{p_{i}}\right)}}
{\frac{1}{2}\log a-2\pi
i\epsilon}\right|_{\stackrel{\epsilon=0}{a=1}}\right\}.
\nonumber
\label{5.3.5}
\end{eqnarray}
A remaining part of the sum~(\ref{5.3.3}) is
\begin{eqnarray}
\frac{1}{2}\sum_{\stackrel{\lambda,m}{\lambda+m+c_{0}\neq 0}}
I(m,\lambda)&=&
e^{2\pi inc_{0}}
\sum_{j=0}^{\infty}\frac{1}{j!}(8\pi iK)^{-j}
\left(\frac{P}{H}\right)^{j}\partial_{\epsilon}^{(2j)}
\left\{e^{2\pi i\epsilon\sum_{i=1}^{n}\frac{\mu_{i}}{p_{i}}}
\right.
\\
&&\left.\left.\times\left[\frac{1}{(2i\sin 2\pi\epsilon)^{n-2}}+
\frac{e^{2\pi i\epsilon(n-4)}}{(n-3)!}\left.\partial_{a}^{(n-3)}
\frac{1}{\frac{1}{2}\log a-2\pi i\epsilon}\right|_{a=1}\right]
\right\}\right|_{\epsilon=0},
\nonumber
\label{5.3.6}
\end{eqnarray}
so that the whole expression for the contribution of a
special reducible connection is
\begin{eqnarray}
Z&=&-\frac{e^{i\frac{\pi}{2}{\rm sign}\left(\frac{H}{P}\right)}}
{\sqrt{2K|H|}}e^{2\pi inc_{0}}
{\rm sign}(P)
\exp-\frac{i\pi}{2K}\left[3{\rm sign}\left(\frac{H}{P}\right)+
\sum_{i=1}^{n}\left(12s(q_{i},p_{i})-\frac{q_{i}}{p_{i}}\right)\right]
\nonumber\\
&&\times\exp 2\pi iK\left(\sum_{i=1}^{n}\frac{r_{i}}{p_{i}}n_{i}^{2}
+\frac{H}{P}c_{0}^{2}\right)
\left\{\frac{1}{2}\sum_{\mu_{1},\ldots,\mu_{n}=\pm 1}
\left[\prod_{i=1}^{n}\mu_{i}\exp\left(2\pi i\mu_{i}
\frac{r_{i}n_{i}+c_{0}}{p_{i}}\right)\right]
\right.\nonumber\\
&&\times\frac{1}{(n-3)!}\int_{0}^{\infty}dx
\left[\left.\partial_{a}^{(n-3)}a^{\frac{1}{2}
\left(x+n-4+\sum_{i=1}^{n}\frac{\mu_{i}}{p_{i}}\right)}
\right|_{a=1}\right]\exp\left(\frac{i\pi}{2K}\frac{P}{H}x^{2}\right)
\nonumber\\
&&+\sum_{j=0}^{\infty}(8\pi iK)^{-j}
\left(\frac{P}{H}\right)^{j}
\\
&&\left.\times
\partial_{\epsilon}^{(2j)}\left.\left[\frac{
\prod_{i=1}^{n}2i\sin\left(2\pi\frac{r_{i}n_{i}+c_{0}+\epsilon}
{p_{i}}\right)}{[2i\sin(2\pi\epsilon)]^{n-2}}+
\frac{1}{(n-3)!}\partial_{a}^{(n-3)}\left.\frac{
a^{\frac{n-4}{2}}\prod_{i=1}^{n}\sin\left(2\pi\frac{
r_{i}n_{i}+c_{0}+\frac{i}{2}\log a}{p_{i}}\right)}
{\frac{1}{2}\log a-2\pi i\epsilon}\right|_{a=1}\right]
\right|_{\epsilon=0}\right\}.
\nonumber
\label{5.3.7}
\end{eqnarray}
It is not hard to see that the term
\begin{equation}
\frac{1}{(n-3)!}\partial_{a}^{(n-3)}\left.\frac{
a^{\frac{n-4}{2}}\prod_{i=1}^{n}\sin\left(2\pi\frac{
r_{i}n_{i}+c_{0}+\frac{i}{2}\log a}{p_{i}}\right)}
{\frac{1}{2}\log a-2\pi i\epsilon}\right|_{a=1}
\label{5.3.8}
\end{equation}
contains only the negative powers of $\epsilon$ in its Laurent series
expansion. Therefore the only purpose of this term is to cancel the
negative powers of $\epsilon$ in the expansion of the term
\begin{equation}
\frac{
\prod_{i=1}^{n}2i\sin\left(2\pi\frac{r_{i}n_{i}+c_{0}+\epsilon}
{p_{i}}\right)}{[2i\sin(2\pi\epsilon)]^{n-2}}
\label{5.3.9}
\end{equation}
so that the whole expression has a smooth limit of
$\epsilon\rightarrow 0$

\noindent
\underline{Trivial Connection}
\nopagebreak

When all $n_{i}=0$, eq.~(\ref{5.3.7}) can be simplified. In
particular, the integral over $x$ and the term~(\ref{5.3.8}) are both
equal to zero after taking a sum over $\mu_{i}$. After adding a
factor~(\ref{3.220}), a total contribution of the trivial connection
is
\begin{eqnarray}
Z&=&-\frac{e^{i\frac{\pi}{2}{\rm sign}\left(\frac{H}{P}\right)}}
{2\sqrt{2K|H|}}e^{2\pi inc_{0}}{\rm sign}(P)
\exp-\frac{i\pi}{2K}\left[3{\rm sign}\left(\frac{H}{P}\right)+
\sum_{i=1}^{n}\left(12s(q_{i},p_{i})-\frac{q_{i}}{p_{i}}\right)\right]
\nonumber\\
&&\times\sum_{j=0}^{\infty}\frac{1}{j!}
\left(\frac{\pi}{2iK}\frac{P}{H}\right)^{j}
\partial_{\epsilon}^{(2j)}
\left.\frac{\prod_{i=1}^{n}2i\sin\left(\frac{\epsilon}{p_{i}}\right)}
{(2i\sin\epsilon)^{n-2}}\right|_{\epsilon=0}.
\label{5.3.10}
\end{eqnarray}

As we have noted in the end of subsection~\ref{*4.3}, a ratio of the
$j=2$ and $j=1$ terms contributes together with the phase $\phi$ of
eq.~(\ref{S.2}) to the 2-loop correction $S_{2}$ as defined by
eq.~(\ref{2.4}). A simple calculation similar to that of
eq.~(\ref{C.1}) shows again that $S_{2}$ is proportional to Casson's
invariant~(\ref{4.1.14}):
\begin{equation}
S_{2}=6\pi\lambda_{CW}.
\label{5.3.11}
\end{equation}
%

%*************************************************
\section{Discussion}
\label{*6}
%*************************************************

In this paper we calculated a full asymptotic large $k$ expansion of
the exact surgery formula for Witten's invariant of Seifert manifolds.
We found a complete agreement between our results and the 1-loop
quantum field theory predictions thus extending the results of the
papers~\cite{FG},\cite{J} on this subject. To achieve this agreement
we had to modify slightly the previous 1-loop formulas for the case
of reducible flat connections and for the case of obstructions in
extending the elements of $H^{1}$ to the moduli of flat connections.

It seems that the method of Poisson resummation used in our
calculations can be applied to Witten's invariants of graph
manifolds, i.e. manifolds constructed by ``plumbing'' the Seifert
manifolds (the solid tori parallel to the fibers are cut out of
Seifert manifolds and the corresponding 2-dimensional boundaries are
glued together after the modular transformations are performed). This
method can also be applied to the invariants built upon simple Lie
groups other than $SU(2)$. We showed that the applicability of the
Poisson resummation is based on
Bernstein-Gelfand-Gelfand resolution.

A rather surprising result of our calculations is the finite loop
exactness of the contributions of irreducible flat connections. This
exactness is somewhat reminiscent of the formulas of paper~\cite{W2}
in which Witten applied a localization principle to the 2-dimensional
gauge theory. The order of the highest loop corrections is equal to
half the dimension of the moduli space of flat connections. This may
suggest that these corrections are related to some intersection
numbers in the moduli space.

The contributions of all flat connections have a specific 2-loop phase
correction $\phi$ (see eq.~(\ref{S.2})). This phase is the same for
all flat connections of a given manifold. In case of a 3-fibered
Seifert manifold, $\phi$ is the only 2-loop correction for the
contribution of irreducible flat connections. The phase $\phi$
looks similar to Casson's invariant, however certain terms are missing
there. It seems, however, that $\phi$ is a manifold invariant in its
own right. It would be interesting to understand its topological
nature.

The ``missing terms'' appear when the full 2-loop correction to the
contribution  of the trivial connection is calculated. The trivial
connection is reducible and its contribution contains an asymptotic
series in $K^{-1}$. The whole 2-loop correction is a combination
of $\phi$ and the second term in this series, and it turns to be
proportional to Casson's invariant. We checked this observation for
$n$-fibered Seifert manifold and found a full agreement with the
formula~(\ref{4.1.14}).

%***********
%here was a piece
%***********

It is worth noting that the change in Casson's invariant under a
surgery on a knot depends on the second derivative of the Alexander
polynomial of that knot (see e.g. \cite{Wa} and references therein).
This derivative is a second order Vassiliev invariant of the knot and
comes as a 2-loop correction to any (i.e., say, either Alexander or
Jones)  knot polynomial if the latter is calculated through Feynman
diagrams. Thus a change in the 2-loop correction under a surgery on a
knot depends on a 2-loop invariant of that knot. We could go a step
backwards and observe a similar relationship between the self-linking
number of a framed knot and the order of homology group (or its
logarithm) of the manifold\footnote{
I am thankful to N. Reshetikhin for discussing the results of his
research on this subject with me.}
(see e.g. \cite{MPR}). Both objects can be interpreted as 1-loop
corrections. It would be interesting to derive a formula (if it
exists) expressing the change in the $n$-loop
correction to the contribution of the trivial connection under a
surgery on a knot through Vassiliev invariants of this knot up
to order $n$.

Let us try to conjecture the formula relating Vassiliev invariants of
a knot to loop corrections of a trivial connection, basing on our
formula~(\ref{S.5}).
The higher loop corrections to the contributions of the reducible
connections (see eqs.~(\ref{S.3}) and (\ref{5.3.4})) are proportional
to the ``derivatives'' of the $U(1)$ Reidemeister torsion. This
looks rather strange in view of the fact that the flat $U(1)$
connections on Seifert manifolds do not have any moduli along which
they could be changed. We therefore propose a different interpretation
of these formulas.

Note that an equation
\begin{equation}
\sum_{j=0}^{\infty}\frac{1}{j!}(8\pi iK)^{-j}
\left(\frac{P}{H}\right)^{j}\left.\partial_{c}^{(2j)}f(c)
\right|_{c=c_{0}}=e^{i\frac{\pi}{4}{\rm sign}\left(\frac{H}{P}
\right)}\sqrt{2K\frac{H}{P}}
\int_{-\infty}^{+\infty}dc\,f(c)\exp\left[-2\pi iK\frac{H}{P}
(c-c_{0})^{2}\right]
\label{6.2}
\end{equation}
can transform the derivatives in eqs.~(\ref{S.3}) and (\ref{5.3.2})
into an integral over $c$. Eq.~(\ref{6.2}) can be derived either by
expanding $f(c)$ in Taylor series at $c=c_{0}$ and checking it for
every term separately, or by noting that the l.h.s. of eq.~(\ref{6.2})
is an exponential of the 1-dimensional Laplacian, which can be
expressed through the heat kernel. In particular, a contribution of
the trivial connection~(\ref{S.5}) can be cast in a form
\begin{equation}
-\frac{e^{i\frac{3}{4}\pi{\rm sign}\left(\frac{H}{P}\right)}}
{2\sqrt{|P|}}{\rm sign}(P)e^{-\frac{i\pi}{2K}\phi}
\int_{-\infty}^{+\infty}d\beta\,e^{-\frac{i\pi}{2K}\frac{H}{P}
\beta^{2}}
\frac{\prod_{i=1}^{3}2i\sin\frac{\pi\beta}{p_{i}}}
{2i\sin\frac{\pi\beta}{K}}.
\label{D.1}
\end{equation}
Here we made a substitution $\beta=2Kc/\pi$.

The formula~(\ref{D.1}) can be given a following interpretation. A
Seifert manifold $X\left(\frac{p_{1}}{q_{1}},\frac{p_{2}}{q_{2}},
\frac{p_{3}}{q_{3}}\right)$ can be constructed by a surgery on a link
consisting of 3 ``fiber'' loops linked to a ``base'' loop (see, e.g.
\cite{FG},\cite{L}). A surgery on the fiber loops produces a connected
sum of three lens spaces. We apply a Reshetikhin-Turaev formula
related to the final surgery on the base loop in order to find
Witten's invariant of $X\left(\frac{p_{1}}{q_{1}},\frac{p_{2}}{q_{2}},
\frac{p_{3}}{q_{3}}\right)$. The Jones polynomial of the base loop
can be expressed as a sum over flat connections in the connected sum
of lens spaces. A contribution of the trivial connection turns out to
be proportional to the integrand of (\ref{D.1}) up to a factor
$\sin(\pi\beta/K)$. Therefore expression~(\ref{D.1}) is actually a
Reshetikhin-Turaev formula in which only a trivial connection
part of the Jones polynomial is taken and a sum over an integrable
weight $\sum_{\beta=1}^{K-1}$ is substituted by an integral
$\frac{1}{2}\int_{-\infty}^{+\infty}d\beta$. We conjecture that these
two changes in the Reshetikhin-Turaev formula produce a
trivial connection contribution to Witten's invariant of the manifold
constructed by a simple $T^{p}S$ surgery on a knot belonging to
some other manifold (in case of a general surgery~(\ref{2.2.2}), only
the $n=0$ term should be retained in the sum of eq.~(\ref{2.3.2})).

Consider now a logarithm of the trivial connection part of Jones
polynomial of a knot. According to \cite{DBN}, the coefficients in its
expansion in powers of $1/K$ are Vassiliev's invariants of the knot. A
1-loop piece in this expansion, which is proportional to
$\pi\beta^{2}/K$, comes from the self-linking number of the knot. This
number can be fractional if the original manifold is nontrivial (for
example, it is equal to $H/P$ in (\ref{D.1})). We conjecture that in
the other terms appearing in the logarithm, the power of $\beta$ is
less or equal to the negative power of $K$. Therefore, if we split off
the self-linking exponential factor and expand the remaining part of
the Jones polynomial in $1/K$, then the gaussian integral in the
modified Reshetikhin-Turaev formula will produce the $1/K$
expansion of the trivial connection contribution to Witten's invariant
of the new manifold. Each term in this expansion will be expressed
through a finite number of Vassiliev invariants of the knot. We hope
to present this calculation in more details in a forthcoming paper.
Here we just want to mention that its results seem to agree \cite{Wa1}
with Walker's formula \cite{Wa} for Casson invariant (if we assume that
Casson invariant is indeed proportional to a 2-loop correction). This
is partly due to the fact that the second derivative of Alexander
polynomial is also a 2-loop correction to Jones polynomial, as
established by D. Bar-Natan in his paper \cite{DBN}.

%**********************************************
\section*{Acknowledgements}
%**********************************************

I want to thank Profs. D.~Auckly, R.~Gompf, C.~Gordon,
N.~Reshetikhin and H.~Saleur for many valuable discussions. I am
especially indebted to Profs. D.~Freed and A.~Vaintrob for their help,
consultations and encouragement.

%**************************************************
\section*{Appendix}
%**************************************************

We present a simple finite dimensional example which illustrates the
appearance of the factor $1/{\rm Vol}(H)$ in the gauge invariant
theory. Consider a 2-dimensional integral
\begin{equation}
I_{\rm gauge}=
\int_{-\infty}^{+\infty}dX_{1}\,\int_{-\infty}^{+\infty}dX_{2}
\exp\left[2\pi i k\,f(\sqrt{X_{1}^{2}+X_{2}^{2}})\right]
\label{A.1}
\end{equation}
for some function $f(r)$. The integrand of this integral is obviously
invariant under the $U(1)$ rotation around the origin. Let us treat
it as a gauge symmetry. Then a ``physical'' quantity would be an
integral $I_{\rm gauge}$ divided by the volume of the gauge group
${\rm Vol}(U(1))=2\pi$. A full machinery of Faddeev-Popov gauge fixing
will lead to a well known expression for the ``physical'' integral:
\begin{equation}
I_{\rm phys}=\frac{I_{\rm gauge}}{{\rm Vol}(U(1))}=
\int_{0}^{\infty}dr\,r\exp(2\pi ikf(r)).
\label{A.2}
\end{equation}
In order to parallel the discussion of subsection~\ref{*2.1} we use a
stationary phase approximation. Suppose that the function $f(r)$ has a
critical point at $r=r_{0}$, i.e. $f^{\prime}(r_{0})=0$. We want to
take an integral~(\ref{A.1}) over the vicinity of that point. We take
a representative point on the gauge orbit:
\begin{equation}
X_{1}=r_{0},\;X_{2}=0.
\label{A.02}
\end{equation}
This is our ``backgroud gauge field''. A simplest choice for the gauge
fixing condition to impose on the fluctuations $(x_{1},x_{2})$
around the background~(\ref{A.02}), would be $x_{2}=0$. However we
make a different choice:
\begin{equation}
kx_{i}v^{i}(X_{0},X_{1})=0,
\label{A.3}
\end{equation}
here $v^{i}$ is a vector field representing the infinitesimal gauge
transformation:
\begin{equation}
v^{i}(X_{1},X_{2})=\epsilon^{ij}X_{j}.
\label{A.4}
\end{equation}
The gauge fixing~(\ref{A.3}) closely resembles a background
covariant gauge fixing $D_{\mu}a_{\mu}$ of the Chern-Simons theory
used in~\cite{W}. Indeed, for an infinitesimal gauge transformation
$\phi$, the analog of $v^{i}$ is $D_{\mu}\phi$ and eq.~(\ref{A.3}) is
similar to a condition
\begin{equation}
\int (D_{\mu}\phi)\;a_{\mu}\;d^{3}x=0
\label{A.04}
\end{equation}
for any function $\phi$, which is equivalent to $D_{\mu}a_{\mu}=0$.

By substituting eqs.~(\ref{A.4}) and~(\ref{A.02}) into eq.~(\ref{A.3})
we get an explicit form of the covariant gauge fixing condition:
\begin{equation}
kx_{2}r_{0}=0.
\label{A.5}
\end{equation}
A Faddeev-Popov ghost determinant is a variation of the gauge fixing
condition with respect to the gauge transformation:
\begin{equation}
\Delta_{\rm gh}=kr_{0}^{2}.
\label{A.6}
\end{equation}
We supplement a quadratic term
$i\pi kf^{\prime\prime}(r_{0})x_{1}^{2}$ in the exponent of
eq.~(\ref{A.1}) with a gauge fixing term $2\pi ikr_{0}yx_{2}$. An
integral over $y$ produces a $\delta$-function of the
condition~(\ref{A.5}). Therefore an operator corresponding to a
quadratic form in the exponent of eq.~(\ref{A.1}) is
\begin{equation}
L_{-}=ik\left(\begin{array}{ccc}
f^{\prime\prime}(r_{0})&0&0\\
0&0&r_{0}\\
0&r_{0}&0\end{array}\right).
\label{A.7}
\end{equation}
The 1-loop ``field-theoretic'' prediction of the physical integral
$I_{\rm phys}$ is
\begin{equation}
I_{\rm phys}=\frac{\det\Delta_{\rm gh}}
{\sqrt{\det(-L_{-})}}e^{2\pi ikf(r_{0})}=
e^{i\frac{pi}{4}}\frac{r_{0}}{\sqrt{kf^{\prime\prime}(r_{0})}}
e^{2\pi ikf(r_{0})}
\label{A.8}
\end{equation}
in full agreement with the stationary phase approximation of the
integral in the r.h.s of eq.~(\ref{A.2}).

Let us see now what happens in the special case when $r_{0}=0$.
A background point~(\ref{A.02}) lies at the origin and is invariant
under the action of $U(1)$. In other words, a ``background field'' has
a $U(1)$ symmetry, so it is similar to a reducible gauge connection
of the Chern-Simons theory. A ghost determinant~(\ref{A.6}) has a zero
mode. The operator $L_{-}$ is different from its ordinary
form~(\ref{A.7}):
\begin{equation}
L_{-}=ik\left(\begin{array}{ccc}
f^{\prime\prime}(0)&0&0\\
0&f^{\prime\prime}(0)&0\\
0&0&0\end{array}\right)
\label{A.9}
\end{equation}
and it also has one zero mode. The prescription for the Reidemeister
torsion~(\ref{2.01}) in a similar situation was to drop the zero
modes. Since no modes are left for the ghosts, we have
\begin{equation}
\Delta_{\rm gh}=1,
\label{A.10}
\end{equation}
while
\begin{equation}
\det(-L_{-})=-\left[kf^{\prime\prime}(0)\right]
\label{A.11}
\end{equation}
A nondegenerate part of the operator~(\ref{A.9}) is the same as if we
were calculating the stationary phase approximation of the
integral~(\ref{A.1}) at the origin without remembering the $U(1)$
symmetry. The same is suggested by the ghost determinant~(\ref{A.10}).
In other words, we see that by dropping the zero modes of ghost
determinant and covariant gauge fixing we ``forgot'' about the $U(1)$
symmetry. Therefore if we substitute expressions~(\ref{A.9}) and
(\ref{A.10}) in the middle part of eq.~(\ref{A.8}), then we will get
the whole integral $I_{\rm gauge}$ rather than its physical ``gauge
fixed'' counterpart $I_{\rm phys}$. So in order to get $I_{\rm
phys}$ from the determinants~(\ref{A.9}) and~(\ref{A.10}) we have to
add a factor $1/{\rm Vol}(U(1))$ ``by hands'' as we did in
eq.~(\ref{2.6}).

\pagebreak
\section*{Figure Captions}
\begin{description}
\item[Fig.1] A fundamental tetrahedron
\item[Fig.2] A section of the fundamental tetrahedron and its Weyl
reflections by a plane $\alpha_{3}={\rm const}$
\item[Fig.3] A section of the full array of terahedra by a plane
$\alpha_{3}={\rm const}$
\item[Fig.4]A fundamental triangle for reducible connections
\item[Fig.5]The zero modes of deformations of a degenerate triangle
\end{description}

\end{document}